\RequirePackage{silence}
\documentclass[trackchanges,twocolumn,tighten]{aastex7}
\newcommand{\pkg}[1]{{\tt #1}}
\shorttitle{\pkg{KilonovaSCORER} for kilonova candidate ranking}
\shortauthors{Darc \& Kilpatrick}
\submitjournal{PASP}
\usepackage{amsmath}
\usepackage{wrapfig}
\usepackage{listings}
\usepackage{tikz}
\usetikzlibrary{arrows.meta,positioning}

\makeatletter
\AtBeginDocument{%
  \@ifundefined{dbltopfraction}{}{%
    \setcounter{dbltopnumber}{4}%
  }%
  \setcounter{topnumber}{8}%
  \setcounter{bottomnumber}{8}%
  \setcounter{totalnumber}{15}%
  \setlength{\textfloatsep}{11pt plus 3pt minus 3pt}%
  \setlength{\floatsep}{9pt plus 2pt minus 2pt}%
  \@ifundefined{dbltextfloatsep}{}{%
    \setlength{\dbltextfloatsep}{11pt plus 3pt minus 3pt}%
  }%
  \@ifundefined{dblfloatsep}{}{%
    \setlength{\dblfloatsep}{9pt plus 2pt minus 2pt}%
  }%
}%
\makeatother

\begin{document}

\title{\pkg{KilonovaSCORER}: Prior-Predictive Scoring of Kilonovae for Real-Time Multimessenger Follow-Up}

\author[0000-0001-8833-474X]{P.~Darc}
\affiliation{Artificial Intelligence for Physics Laboratory (Lab-IA), \\ 
Centro Brasileiro de Pesquisas F\'{i}sicas, Rua Dr.\ Xavier Sigaud 150, CEP 22290-180, Rio de Janeiro, RJ, Brazil}
\affiliation{Center for Interdisciplinary Exploration and Research in Astrophysics (CIERA), \\
Northwestern University, Evanston, IL 60201, USA}
\email{phelipedarc@gmail.com}

\author[0000-0002-5740-7747]{C. D. Kilpatrick}
\affiliation{Center for Interdisciplinary Exploration and Research in Astrophysics (CIERA), \\
Northwestern University, Evanston, IL 60201, USA}
\email{ckilpatrick@northwestern.edu}

\correspondingauthor{P. Darc}
\email{phelipedarc@cbpf.br}

\begin{abstract}
Real-time ranking of optical transient candidates during gravitational-wave (GW) and multimessenger follow-up is challenging when only sparse early-time, multi-band photometry is available.
We present \pkg{KilonovaSCORER}, an open-source framework for scoring and ranking in this regime.
It quantifies the consistency of each candidate with a physically
motivated kilonova model grid in absolute magnitude space using two
complementary per-observation metrics, $P_{\mathrm{tail},\mathrm{KNe}}$
and $P_{\mathrm{near},\mathrm{KNe}}$.
These are aggregated into a cumulative ranking score via inverse-variance weighting in logit
space, naturally accounting for heterogeneous observational uncertainties across bands and epochs.
A sequential Approximate Bayesian Computation (ABC) diagnostic tracks
photometric consistency across epochs, penalizing candidates
whose temporal evolution is incompatible with kilonova expectations.
We validate the framework on AT\,2017gfo and SN\,2025ulz, and test
it against supernova simulations under a realistic Rubin/LSST
Target-of-Opportunity strategy. The framework recovers kilonova
candidates with high confidence while ruling out supernova contaminants within five days of the gravitational-wave trigger.
In our LSST ToO simulations, median cumulative scores for thermonuclear and core-collapse supernova contaminants fall to zero by $3$--$4$\,d post-trigger, whereas kilonova medians remain $\gtrsim 0.4$.
\pkg{KilonovaSCORER} supports real-time workflows for ToO teams and LSST alert brokers, integrates with follow-up coordination platforms such as the Tool for Rapid Object Vetting and
Examination, and is publicly available at \url{https://github.com/phelipedarc/KilonovaSCORER/tree/main}.
\end{abstract}

\section{Introduction}
The Vera C. Rubin Observatory Legacy Survey of Space and Time (LSST) is designed to survey the southern sky deeper and faster than any previous wide-field optical survey \citep{verarubin2019ApJ...873..111I}. LSST will enable the discovery of an unprecedented number of astrophysical transients, opening a new era of large-scale optical time-domain surveys \citep{fink10.1093/mnras/staa3602}. Southern-sky wide-field imaging is likewise being extended by the La Silla Southern Schmidt Survey (LS4; \citep{LS4_Miller2025}). About $10^{7}$ alerts per night are expected to be ingested by the alert brokers \citep{fink10.1093/mnras/staa3602,verarubin2019ApJ...873..111I,LASAIR_Smith2019,Alerce_Frster2021,babamul_booom_coughlin,Antares_matheson,Bellm2019PASP} and subsequently distributed to the community through web portals and application programming interfaces (APIs). A fraction of these alerts will be temporally and spatially
coincident with gravitational-wave events \citep{abbott2020prospects,Abbott2017GW170817GW}, gamma-ray bursts (GRBs; GRB for a single burst) \citep{CTA_gammaray,RastinajGRB211211A,SVOMgammaandXrays}, and high-energy neutrino detections \citep{neutrino_keck_AdrinMartnez2016}. Consequently, LSST has the potential to rapidly detect and identify electromagnetic (EM) counterparts to multimessenger sources, such as kilonovae (KNe; KN for a single kilonova), particularly at large distances where typical counterparts ($M \sim -16$ mag in the optical) are expected to be too faint for most existing wide-field survey telescopes. 

Kilonovae serve as key probes of heavy-element production in astrophysical environments \citep{Roederer2018}, and provide independent constraints on the neutron star equation of state \citep{Radice_2018,Abbott2017GW170817GW} as well as on the expansion rate of the Universe through multimessenger observations \citep{hubble_kn}. Joint modeling of gravitational-wave and electromagnetic data is an active area of development \citep{Breschi2024AandA689,Pang2023NatCommNMMA}.

Future kilonova detections will mostly rely on rapid target-of-opportunity (ToO) observations \citep{too_Andreoni2022}, which are expected to require an investment of $\sim$2\% of the total survey time budget (corresponding to $\sim$96.2 hours per LSST year).\footnote{The ToO triggering strategy is designed to efficiently follow up well-localized gravitational-wave events. Observations are initiated for events whose 90\% credible sky localization region subtends fewer than $\sim500$\,deg$^{2}$ and whose merger classification yields a probability greater than 90\% of being a binary neutron star (BNS) or neutron star--black hole (NSBH) system. Events satisfying these criteria are divided into two tiers based on localization quality and signal significance. Under the \textit{gold} strategy, observations are conducted over the first 4 days following the merger trigger: multi-band imaging in three filters is obtained on the first night, followed by two-filter coverage on each subsequent night. The \textit{silver} strategy follows the same temporal cadence but employs two-filter imaging throughout all four nights.}

Multimessenger astronomy is evolving rapidly, and rapid vetting and classification of the immense number of transient candidates is essential \citep{Kilpatrick_2021_gravcole,Rastinejad2022O3KNcandidates,gravicoulter2024gravitycollectivecomprehensiveanalysis}. This challenge is particularly critical for transients lying within the high-probability localization regions of gravitational-wave (GW) events. Such regions can span tens to hundreds of square degrees and encompass a substantial background of unrelated astrophysical variability and explosive phenomena. Among this large number of candidates, identifying the true electromagnetic counterpart requires physics-informed vetting of every candidate based on the available data. 

Kilonovae are powered by the radioactive decay of heavy elements synthesized through rapid neutron-capture (r-process) nucleosynthesis \citep{Li_1998}. Observationally, kilonovae are fast-evolving ultraviolet--optical--infrared transients, with light curves that typically peak $\simeq$ 2--3 days after merger \citep{Metzger_2019}. Consequently, the window for detailed characterization of the early-time ejecta closes within $\lesssim 4$\,d of merger. Allocating spectroscopic and multi-wavelength follow-up resources to the wrong candidates during this window is irreversible. Moreover, limited coordination of follow-up observations compounds the problem and hinders efficient identification of true multimessenger counterparts.

Most wide-field detections exhibit poorly sampled light curves, often with only a single high-significance epoch. As shown by \citet{Anaistevenson2026}, simulations generated under the standard LSST survey strategy using two widely adopted kilonova light-curve models (the Kasen model \citep{Kasen_2017}, used in the Photometric LSST Astronomical Time-series Classification Challenge (PLAsTiCC) \citep{Plasticc_Kessler2019}, and the Bulla model \citep{bulla2019}, employed in the ELAsTiCC classification challenge) demonstrate the observational challenges inherent to kilonova detection. Only $\sim 11\%$ (Bulla) and $\sim 7\%$ (Kasen) of simulated light curves have at least one high signal-to-noise ratio (SNR) detection that would be sent to the brokers, and fewer than $\sim 6\%$ of the generated light curves produce at least two detections with $\mathrm{SNR} \geq 5$.

Similarly, in a ToO scenario, we expect most candidates observed during the first few nights to have fewer than four observations. In that regime, full Bayesian parameter inference and model comparison are often weakly informative and sensitive to prior assumptions \citep{Raaijmakers2021ApJ922}. Machine learning classifiers developed for LSST brokers \citep{Fraga2024, Alerce_Frster2021} are primarily optimized for large-scale taxonomic classification under regular survey cadence and are not designed for the low-data, early-time regime of GW follow-up. Consequently, it is essential to develop tools capable of scoring potential kilonova candidates using only a sparse set of early-time observations, without requiring full posterior inference for every alert. These tools support rapid decision-making and guide the prioritization of follow-up campaigns, particularly during the fast-rising phase of the transient, when timely observations are crucial for probing the underlying physical processes in detail.

In this work, we introduce \pkg{KilonovaSCORER}, a simulation-based statistical framework developed to score and rank transient candidates according to their consistency with kilonova light-curve models in absolute-magnitude space. In short, it evaluates two complementary per-observation metrics,
$P_{\mathrm{tail},\mathrm{KNe}}$ and $P_{\mathrm{near},\mathrm{KNe}}$, which together quantify the local and global consistency of each observation with the kilonova prior predictive distribution. Individual scores are
aggregated into a cumulative ranking score via inverse-variance weighting in logit space, with a score uncertainty that narrows as additional observations arrive. A complementary Approximate Bayesian
Computation (ABC) sequential diagnostic tracks the global consistency of the candidate's photometric history across all epochs, immediately penalizing candidates whose temporal evolution becomes incompatible
with any model in the kilonova grid. Together, these components produce a final score with propagated uncertainty and a physically interpretable consistency flag, updated in real time as new photometry is ingested.

\pkg{KilonovaSCORER} is designed to integrate directly with \pkg{TROVE} (Tool for Rapid Object Vetting and Examination, \citealt{Franz2025, nick_trove})\footnote{\url{https://astro-trove.github.io/}},
an open-source infrastructure for real-time candidate ranking that ingests
data from LSST alert brokers, the Transient Name Server (TNS), and
the General Coordinates Network (GCN). Within \pkg{TROVE}, \pkg{KilonovaSCORER} acts as a dedicated physics-informed photometric scoring layer, complementing the existing 3D localization, catalog cross-match, and aggregate photometric metrics.

To evaluate the robustness of \pkg{KilonovaSCORER}, we subject the framework to four validation cases. First, we score AT\,2017gfo, the gold-standard and only confirmed optical counterpart
to a GW event, to establish a performance baseline. Second, we analyze SN\,2025ulz, a Type IIb supernova with a kilonova-like early evolution, to test the Approximate Bayesian Computation (ABC) sequential diagnostic's ability to rule out false positives during follow-up. Third, we apply \pkg{KilonovaSCORER} to a sample of GRB-associated kilonovae to assess consistency across diverse viewing geometries and observational contexts. Finally, we use realistic Rubin/LSST ToO simulations (incorporating BNS, NSBH, and three supernova emission models) to statistically evaluate the cumulative score evolution for each transient and its ranking capabilities within the first four nights following a gravitational-wave trigger.

This paper is organized as follows. In Section~\ref{sec:data}, we describe the kilonova simulations adopted in this work, together with the real observational data from previous follow-up campaigns used to validate our methodology. Section~\ref{sec:methods} presents the statistical framework underlying the scoring system, including the definition of the scoring metrics and the model-consistency diagnostics. In Section~\ref{sec:results}, we evaluate the performance of \pkg{KilonovaSCORER} using both real candidates and simulated transient events. Section~\ref{sec:discussion} discusses the limitations of the approach and its applicability to realistic follow-up scenarios. Finally, Section~\ref{sec:software_data} summarizes software access and the data sources used for validation.

\section{Data}\label{sec:data}

\subsection{SN\,2025ulz and Candidate Counterparts}

The GW event S250818k was initially reported by the LIGO/Virgo/KAGRA (LVK) Collaboration as a sub-threshold trigger, indicating that at least one component of the system was consistent with a sub-solar–mass black hole or neutron star. The event was assigned a 29\% probability of originating from a BNS merger and a 71\% probability of being of terrestrial origin. This initial classification yielded a 50\% (90\%) credible sky localization region of 205\,deg$^{2}$ (786\,deg$^{2}$), together with a posterior luminosity distance estimate of $259 \pm 62$\,Mpc. Despite the relatively low astrophysical probability, S250818k attracted significant interest for several reasons. First, it was among the few candidates during the fourth observing run of LVK with a non-zero probability of being a BNS merger. Second, an optical transient (SN\,2025ulz) was discovered within the localization region, sparking extensive discussion regarding a potential association with the GW trigger \citep{ulz_Kasliwal2025,ulz_OConnor2025,ulz_panstars,Franz2025}.

During the first few days following discovery ($\lesssim 2.5$ days), SN\,2025ulz exhibited a rapid photometric decline, progressive color reddening, and featureless optical spectra, properties broadly consistent with expectations for kilonova-like transients. However, subsequent follow-up observations obtained at later epochs ($\sim$5 days post-discovery) revealed a flattening and subsequent rebrightening of the light curve. These behaviors are inconsistent with kilonova evolution and instead favor an interpretation as a Type~IIb supernova, as discussed by \citet{Franz2025}.

We retrieved all publicly available observations through the \pkg{TROVE} system, compiling data initially reported via the GCN and TNS for SN\,2025ulz. The dataset includes ultraviolet, optical, infrared, and radio measurements reported in GCN circulars, together with additional follow-up observations presented by \citet{Franz2025}. This event provides an ideal test case for validating our kilonova scoring framework under a realistic BNS search scenario.

\subsection{GW170817: AT\,2017gfo}

To date, GW170817 remains the first and only confirmed joint gravitational-wave and electromagnetic detection, observed in association with a short gamma-ray burst (\citealt{abbott2017search}), the optical kilonova AT\,2017gfo (e.g., \citealt{Coulter17,Soaressantos2017ApJ...848L..16S}), and long-lived radio (e.g., \citealt{2017alex}) and X-ray (e.g., \citealt{17gfo_Margutti2017}) afterglow emission. We refer the reader to \citet{Margutti2021} for a comprehensive review of the gravitational-wave and electromagnetic observations of this event. The GW170817 event was reported with a posterior luminosity distance of $38.58 \pm 6.99$\,Mpc. The associated optical counterpart, AT\,2017gfo, was localized to a host galaxy at a redshift of $z \approx 0.0098$. For this study, we compiled all publicly available multi-band photometric observations of AT\,2017gfo from the Open Supernova Catalog through the \pkg{redback} API \citep{REDBACKSarin2024}.

\begin{table*}[t]
\centering 
\caption{Prior distributions adopted for the two-component kilonova model.}
\label{tab:priors_2comp_kne}
\begin{tabular}{lcc}
\hline
\hline
Parameter & Prior range & Description \\
\hline

Ejecta mass $M_{\mathrm{ej,[1,2]}}$ 
& $\mathcal{U}(10^{-4},\;0.1)\,M_{\odot}$ 
& Ejecta mass of each component \\

Ejecta velocity $v_{\mathrm{ej,[1,2]}}$ 
& $\mathcal{U}(0.01,\;0.7)\,c$ 
& Ejecta velocity of each component \\

Opacity $\kappa_{1}$ 
& $\mathcal{U}(0.01,\;0.5)\,\mathrm{cm}^{2}\,\mathrm{g}^{-1}$ 
& Blue-component opacity \\

Opacity $\kappa_{2}$ 
& $\mathcal{U}(1,\;30)\,\mathrm{cm}^{2}\,\mathrm{g}^{-1}$ 
& Red-component opacity \\

Temperature floor $T_{\mathrm{floor,[1,2]}}$ 
& $\log\mathcal{U}(100,\;6000)\,\mathrm{K}$ 
& Temperature floor of each component \\

\hline
\hline
\end{tabular}
\end{table*}

\subsection{Kilonova Simulations}

Kilonova light curves are mainly characterized by the physical properties of the merger ejecta, including the ejecta mass ($M_{\mathrm{ej}}$), ejecta velocity ($v_{\mathrm{ej}}$), and opacity ($\kappa$) \citep{Kasen_2017,Villar17,Metzger_2019}. Early radiative transfer simulations indicate that kilonova emission can generally be described by two distinct components. The ``red'' component is dominated by near-infrared (NIR) emission arising from lanthanide-rich ejecta with high opacity ($\kappa \gtrsim 1\,\mathrm{cm^{2}\,g^{-1}}$). The second is a ``blue'' component produced by ejecta containing lighter $r$-process elements, characterized by lower opacity ($\kappa \sim 0.5\,\mathrm{cm^{2}\,g^{-1}}$).

\pkg{KilonovaSCORER} adopts simulations from the two-component kilonova model presented by \citet{Villar17}, since one-component kilonova models are unable to fully reproduce the early blue emission observed in AT\,2017gfo. This model is parameterized, for each component, by the ejecta mass ($m_{\mathrm{ej}}$, in units of solar masses), the minimum initial velocity ($v_{\mathrm{ej}}$, expressed in units of the speed of light), the floor temperature ($T_{floor}$, in Kelvin), and the ejecta opacity ($\kappa_{\mathrm{ej}}$, in $\mathrm{cm^{2}\,g^{-1}}$), resulting in a total of eight free parameters. A key assumption of this model is that the two emission components are modeled independently. The total flux at each wavelength is computed as the linear sum of the two components, and radiative reprocessing between the ejecta components is not included.

Because our goal is to score potential kilonova candidates in a way that remains as model-independent as possible, we adopt broader prior distributions than those typically used for detailed kilonova parameter inference studies. This choice is designed to encompass a wide range of plausible kilonova-like behaviors, thereby maximizing sensitivity to genuine kilonovae while enabling the rejection of transients that are inconsistent with this general class of models. The adopted prior distributions, $p(\theta)$, for all model parameters are summarized in Table~\ref{tab:priors_2comp_kne}. Two aspects of our prior choices are particularly important. First, unlike \citet{2comp_Villar2017}, who investigated both two- and three-component models with the blue-component opacity fixed at $\kappa = 0.5\,\mathrm{cm^{2}\,g^{-1}}$, we allow this parameter to vary over the range $0.01 \leq \kappa \leq 0.5\,\mathrm{cm^{2}\,g^{-1}}$. Second, as discussed by \citet{kne_Kitamura2025}, the widely used “red” and “blue” components in analytic kilonova models should not necessarily be interpreted as direct representations of physically distinct ejecta components, such as dynamical ejecta and post-merger winds, respectively. Consequently, we adopt the broader parameter space suggested by \citet{kne_Kitamura2025}, allowing relatively extreme parameter values, including ejecta masses as low as $10^{-4}\,M_\odot$ and ejecta velocities up to $0.7c$ for both the red and blue components. This broader interval enables exploration of a richer diversity of expected emission behaviors from binary neutron star mergers, while avoiding overly restrictive physical assumptions during the candidate-scoring stage.

\begin{figure*}[!htbp]
    \centering
    \includegraphics[width=\textwidth]{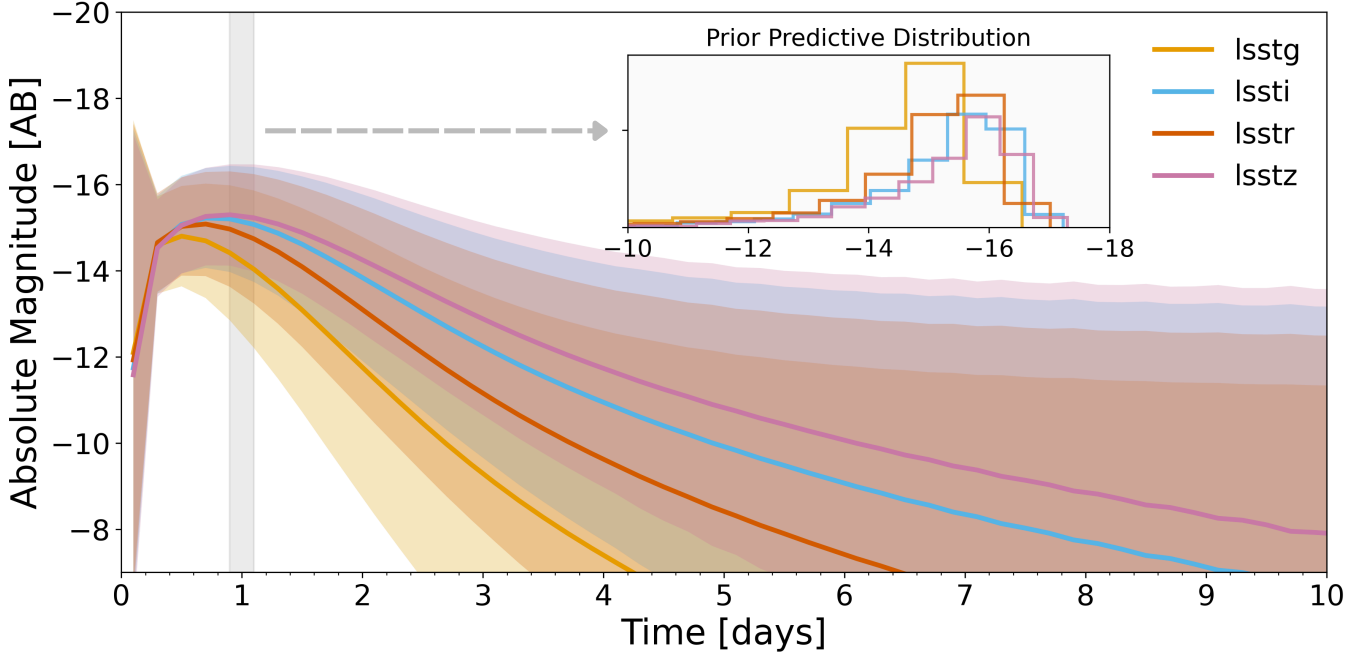}
    \caption{Simulated multi-band kilonova light curves from the
    adopted two-component model grid, showing the evolution of the
    prior predictive distribution in absolute magnitude over the first
    10\,d after merger. Solid lines show the median magnitude in each
    photometric band within 0.2\,d time bins, and shaded regions
    indicate the corresponding $1\sigma$ spread across all prior
    samples. The inset shows the absolute magnitude distribution at
    $t = 1$\,d (shaded grey band, $\Delta t = 0.2$\,d), illustrating
    the prior predictive distribution discussed in Section~\ref{sec:ppd}.
    Horizontal axis: time since merger (d). Vertical axis: absolute magnitude $M$ in each LSST band (AB).}
    \label{fig:simulated_lcs}
\end{figure*}

\textbf{Simulation Setup:} To construct the reference distribution (hereafter Prior Predictive Distribution, Section~\ref{sec:kilonovascorer}) against which candidate observations are compared, we simulate $N_{\rm sim} = 10^{5}$ kilonova light curves in the LSST $g$, $r$, $i$, and $z$ bands. Parameter sets $\theta^{*}$ are drawn from the prior $p(\theta)$ and passed through the two-component radiative transfer model implemented in \pkg{redback} \citep{REDBACKSarin2024}. The resulting ensemble of simulated light curves in absolute magnitude space is shown in Figure~\ref{fig:simulated_lcs}. Each light curve spans the first 10 days post-merger, consistent with the typical timescale of kilonova emission, and is evaluated on a uniform grid of $10^{3}$ time steps. Observations are grouped into time bins of width $\Delta t 
= 0.2$\,d, chosen to be representative of the typical separation between exposures obtained within a single LSST observing night. Each bin therefore contains ${\sim}20$ simulated magnitude samples per band per parameter realization, yielding a total of $20 \times N_{\mathrm{sim}}$ model-predicted absolute-magnitude samples per time bin per band. This ensemble is sufficiently large to robustly approximate the prior predictive distribution $p(M \mid t, \mathrm{band})$ against which individual photometric detections are 
scored. 

Several binning strategies can be implemented within this framework. To optimize computational speed and memory usage, simulations may be pre-binned and stored as chunks of magnitude distributions for each band and time interval, loading only the subsets corresponding to the observation epochs. Alternatively, in a computationally resource-rich setup, binning may be performed directly around the observation times, ensuring that simulations cover a temporal window extending 0.1 MJD before and after each measurement.\footnote{We find that the choice of binning strategy has a negligible impact on the resulting scores. Consequently, we adopt a scheme in which the first and last observations fall at the center of their respective time bins. This strategy balances temporal resolution and computational efficiency, enabling reliable and scalable candidate scoring within an alert-driven framework. Future versions of \pkg{KilonovaSCORER} will include additional binning strategies, providing greater flexibility to adapt the scoring procedure to different computational environments.}

\subsection{Uncertainty Quantification \& Preprocessing}

\textbf{Observations:} The scoring methodology is specifically designed to assess and rank transient candidates identified through either LSST alerts or initial observations reported to the TNS. The primary objective is to prioritize candidates exhibiting the highest likelihood of being a kilonova-like transient. Accordingly, the information used for each observation is limited to the apparent magnitude $m_i$, the corresponding photometric uncertainty $\sigma_{m,i}$, the observation time relative to the merger epoch $t_{obs,i}$, and the photometric filter ($i$). In addition, for each candidate set we incorporate information from the associated gravitational-wave event, specifically the inferred luminosity distance and its uncertainty.

\textbf{Data uncertainty \& preprocessing:} Uncertainties are propagated 
through a Monte Carlo sampling scheme that accounts for two principal sources 
of uncertainty: photometric measurement uncertainty and luminosity distance 
uncertainty. For each candidate, we require that it lies within the 
gravitational-wave localization volume, so the luminosity distance is therefore 
sampled from a Gaussian distribution $\mathcal{N}(D_L, \sigma_{D_L})$, where 
$D_L$ and $\sigma_{D_L}$ are the posterior mean and standard deviation reported 
by the GW parameter estimation pipeline. Independently, the observed apparent 
magnitude is sampled from $\mathcal{N}(m_{\mathrm{app}}, \sigma_m)$, where 
$\sigma_m$ is the photometric uncertainty of the detection. This joint sampling framework enables a consistent propagation of uncertainties arising from both photometric measurement errors and luminosity distance uncertainty when converting apparent magnitudes to absolute magnitudes via the distance modulus. As a result, each observation is associated with a Region of Practical Equivalence (ROPE), defined as an interval in absolute-magnitude space within which all sampled values are considered statistically equivalent.

\section{Methods} \label{sec:methods}
\subsection{\pkg{TROVE} Infrastructure} \label{sec:trove}

\pkg{TROVE} (Tool for Rapid Object Vetting and Examination,
\citealt{Franz2025, nick_trove})\footnote{\url{https://astro-trove.github.io/}}
is an open-source, API-enabled platform for the real-time evaluation
and ranking of transient candidates associated with multimessenger
events, including gravitational-wave triggers, gamma-ray bursts, and
high-energy neutrino alerts. \pkg{TROVE} ingests observational data from
LSST alert brokers, the Transient Name Server (TNS), and the General
Coordinates Network (GCN), evaluating each candidate against a suite
of complementary metrics: a three-dimensional localization score
quantifying spatial consistency with the gravitational-wave sky
volume, a point-source and artifact score cross-matching candidates
against public catalogs of known variable stars, active galactic nuclei (AGN), and asteroids
within $2''$, and a photometric score. The final \pkg{TROVE} score is the product of all individual
metrics, ensuring a candidate must satisfy all criteria simultaneously
to receive a high overall ranking. We refer the reader to
\citet{Franz2025} and \citet{nick_trove} for a full description of
the implementation.

During a real follow-up campaign, tens to hundreds of candidates may be ingested within the first night, each requiring rapid photometric assessment before spectroscopic resources can be allocated. 
\pkg{KilonovaSCORER} is designed to operate directly on the estimated $M_{\mathrm{obs}},\sigma_{\mathrm{obs}}$ of each candidate as ingested in real time by \pkg{TROVE}. It acts as a new simulation-based photometric scoring layer, enabling physics-informed kilonova vetting at the scale and cadence demanded by
Rubin/LSST gravitational-wave follow-up campaigns.

\subsection{\texorpdfstring{\pkg{KilonovaSCORER}: Statistical Framework for the Scoring System}{KilonovaSCORER: Statistical Framework for the Scoring System}}\label{sec:kilonovascorer}

\pkg{KilonovaSCORER} is formulated within a Bayesian framework, 
based on ideas from prior 
predictive checking (PPC) \citep{ppcBox1980, ppcGelman1996} and 
simulation-based calibration (SBC) \citep{talts2020validating, 
priorSBC_Modrk2025}. Prior predictive checking is traditionally used to assess whether a prior is appropriate by generating synthetic data before conditioning on observations. SBC is used to validate sampling algorithms such as Markov chain Monte Carlo (MCMC), nested sampling, or other numerical Bayesian inference methods. We adapt these concepts to a different purpose. Rather than using them as diagnostics for inference algorithms, we use the prior predictive distribution directly as a scoring reference, quantifying whether a candidate's photometric observations constitute a plausible realization of what the kilonova model considers possible \citep{Gabry2019}.

The kilonova simulation grid serves as a Monte Carlo representation of this 
prior predictive distribution: at a given epoch and rest-frame band, the 
ensemble of simulated light curves defines the range of absolute magnitudes 
implied by the physical model and its parameter priors, without being tuned to any specific observation. Candidate apparent magnitudes are converted to absolute magnitude space using the GW-inferred luminosity distance and its associated uncertainty (Section~\ref{sec:data}), and each observation is evaluated for consistency with this distribution via two complementary statistical scores. By never conditioning on the observed data, this approach is inherently conservative, guarding against overfitting and artificially optimistic model agreement, which is particularly valuable at early times when sparse photometry cannot support reliable inference across a high-dimensional kilonova parameter space.

\begin{figure*}[t!]
    \centering
    \includegraphics[width=1\linewidth]{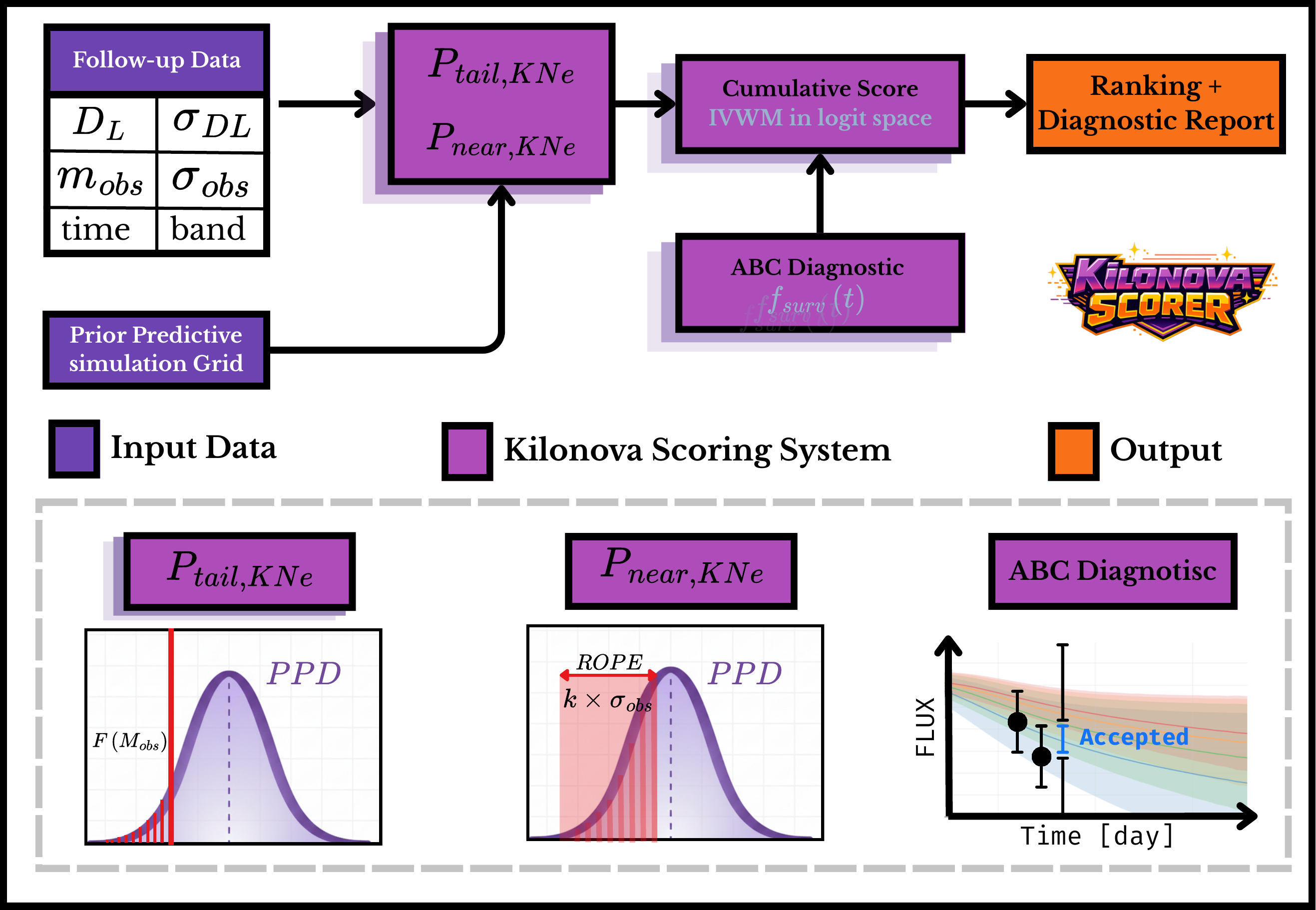}
    \caption{Schematic data flow of the \pkg{KilonovaSCORER} framework. The pipeline ingests multi-band photometry and gravitational-wave distance estimates (indigo boxes), scoring candidates by comparing observations against a prior predictive simulation grid in absolute-magnitude space (purple boxes). Three complementary metrics (illustrated in the grey dashed boxes) quantify kilonova consistency: \textit{Left panel:} $P_{\mathrm{tail},\mathrm{KNe}}$ is computed from the cumulative distribution function $F(M_{\mathrm{obs}})$ of the prior predictive distribution at each epoch and band, with observational uncertainties propagated via $N$ samples from $\mathcal{N}(M_{\mathrm{obs}}, \sigma_{\mathrm{obs}})$. Individual epoch scores are combined into a cumulative ranking score via inverse-variance weighting in logit space. \textit{Middle panel:} $P_{\mathrm{near},\mathrm{KNe}}$ assesses local consistency by measuring the fraction of simulations within a Region of Practical Equivalence (ROPE) defined by user parameter $k_{\mathrm{near}}$. \textit{Right panel:} The Approximate Bayesian Computation (ABC) diagnostic tracks temporal evolution through a survival fraction $f_{\mathrm{surv}}(t)$ of kilonova models satisfying successive ROPE criteria, defined by user parameter $k_{\mathrm{ABC}}$. When candidate evolution becomes physically incompatible with the model grid, the survival fraction and cumulative score collapse to zero, enabling early rejection of non-kilonova transients. The cumulative score with associated uncertainty ranks gravitational-wave follow-up candidates, while the $P_{\mathrm{near},\mathrm{KNe}}$ and ABC diagnostics are compiled into a vetting report for rapid candidate assessment (Orange Box).}
    \label{fig:pipeline_score}
\end{figure*}

Throughout this section, we describe the construction of the prior predictive distribution, the definition of the two implemented ranking scores 
($P_{\mathrm{tail},\mathrm{KNe}}$ and $P_{\mathrm{near},\mathrm{KNe}}$), and the Approximate 
Bayesian Computation (ABC) diagnostic used to evaluate the temporal consistency of candidate light curves. Figure~\ref{fig:pipeline_score} summarizes the end-to-end flow from inputs to ranked output, and the following subsections define each stage formally.

\subsection{Prior Predictive Distribution}\label{sec:ppd}

As discussed in Section~\ref{sec:data}, the prior predictive distribution (PPD) used in our scoring methodology is constructed by generating a population of synthetic kilonova light curves. Specifically, we simulate $N_{\rm sim}$ light curves by drawing model parameters $\boldsymbol{\theta}^{(i)}$ from the prior distribution $p(\boldsymbol{\theta})$ and evaluating the kilonova model for each sampled parameter set.

For a given observation obtained at time $t$ and in photometric filter $b$, we extract the corresponding absolute magnitudes from all simulated light curves within a time window of $\pm 0.2$ days around $t$. The resulting ensemble of model-predicted absolute magnitudes, denoted $M_{\mathrm{model}}$, defines a Monte Carlo representation of the prior predictive distribution of the absolute magnitude at $(t,b)$.

Formally, the prior predictive distribution is defined as
\begin{equation}
p\!\left(M_{\mathrm{model}} \mid t, b \right)
=
\int
p\!\left(M_{\mathrm{model}} \mid \boldsymbol{\theta}, t, b \right)\,
p(\boldsymbol{\theta})\,
d\boldsymbol{\theta},
\end{equation}
where $p(\boldsymbol{\theta})$ is the prior distribution over the model parameters, and 
$p\!\left(M_{\mathrm{model}} \mid \boldsymbol{\theta}, t, b \right)$ is the likelihood implied by the forward kilonova model, i.e., the probability density of obtaining an absolute magnitude $M_{\mathrm{model}}$ at time $t$ in band $b$ given a specific parameter set $\boldsymbol{\theta}$.

In practice, this multidimensional integral is not evaluated analytically. Instead, it is approximated via Monte Carlo sampling,
\begin{equation}
p\!\left(M_{\mathrm{model}} \mid t, b \right)
\approx
\frac{1}{N_{\rm sim}}
\sum_{i=1}^{N_{\rm sim}}
p\!\left(M_{\mathrm{model}} \mid \boldsymbol{\theta}^{(i)}, t, b \right),
\end{equation}
where $\boldsymbol{\theta}^{(i)} \sim p(\boldsymbol{\theta})$. 

This distribution represents the range of magnitudes that the model predicts as physically plausible under the assumed priors on the model parameters. At this stage, the PPD reflects only the intrinsic variability of the model and does not yet account for observational uncertainties. To enable a conservative comparison with the data, we incorporate the uncertainty associated with the observed absolute magnitude. For a single observation with absolute magnitude $M_{\mathrm{obs}}$ and uncertainty $\sigma_{\mathrm{obs}}$ \footnote{derived from the photometric uncertainty $\sigma_{m_{\mathrm{obs}}}$ and the luminosity distance uncertainty $\sigma_{D_L}$}, we generate a replicated data point $M_{\rm rep}$ by adding Gaussian white noise to the model predictions,
\begin{equation}
M_{\mathrm{rep}} = M_{\mathrm{model}} + \epsilon, \qquad
\epsilon \sim \mathcal{N}(0, \sigma_{\mathrm{obs}}^{2}).
\end{equation}

Conceptually, this is equivalent to the convolution
\begin{equation}
\begin{split}
p\!\left(M_{\mathrm{rep}} \mid t, b \right)
&=
\int
p\!\left(M_{\mathrm{rep}} \mid M_{\mathrm{model}}, \sigma_{\mathrm{obs}} \right)\,
\\
&\quad {}\times
p\!\left(M_{\mathrm{model}} \mid t, b \right)
\, d M_{\mathrm{model}},
\end{split}
\end{equation}
where the observational noise is modeled as a Gaussian,
\begin{equation}
\begin{split}
p\!\left(M_{\mathrm{rep}} \mid M_{\mathrm{model}}, \sigma_{\mathrm{obs}} \right)
={} &
\frac{1}{\sqrt{2\pi}\,\sigma_{\mathrm{obs}}} \\
& \times \exp\!\left[-\frac{(M_{\mathrm{rep}} - M_{\mathrm{model}})^2}{2 \sigma_{\mathrm{obs}}^2}\right].
\end{split}
\end{equation}

This procedure yields a noise-convolved prior predictive distribution that represents the magnitudes we would expect to measure given our current understanding of kilonova physics and the observational uncertainty. This extra step of convolution broadens the prior predictive distribution in accordance with $\sigma_{\mathrm{obs}}$, ensuring that both intrinsic model variability and observational uncertainty are taken into account, leading to a more conservative PPD.

In this framework, the comparison is between two objects: (i) the noise-convolved prior predictive distribution $p\!\left(M_{\mathrm{rep}} \mid t, b \right)$, which encodes the model’s expectations for absolute magnitudes, and (ii) the observed data, represented by a single measurement $M_{\mathrm{obs}}$, which can be conceptualized as a Gaussian distribution centered at $M_{\mathrm{obs}}$ with standard deviation $\sigma_{\mathrm{obs}}$.

\subsection{Prior Predictive Check \& Model Consistency}

Given the noise-convolved prior predictive distribution of replicated magnitudes $p\!\left(M_{\mathrm{rep}} \mid t, b \right)$ and an observed absolute magnitude $M_{\mathrm{obs}}$ with uncertainty $\sigma_{\mathrm{obs}}$, we define model consistency as the degree to which the observed data are typical of realizations generated by the model.

\subsubsection{\texorpdfstring{$P_{\mathrm{tail},\mathrm{KNe}}$ Score}{P-tail KNe Score}}

To quantify how extreme a measured absolute magnitude is relative to the noise-convolved prior predictive distribution of replicated magnitudes, $p(M_{\mathrm{rep}} \mid t, b)$, we define a score based on a two-sided tail-area probability. This quantity measures the degree to which the observation lies in the tails of the PPD, serving as a scalar indicator of model-data consistency.

Conceptually, the $P_{\mathrm{tail},\mathrm{KNe}}$ score can be expressed in terms of the cumulative distribution function (CDF) of the prior predictive distribution:

\begin{equation}
F(M_{\mathrm{obs}}) = \int_{-\infty}^{M_{\mathrm{obs}}} p(M_{\mathrm{rep}} \mid t, b)\, dM_{\mathrm{rep}} \, ,
\end{equation}

which corresponds to the cumulative probability, under the model, of obtaining a magnitude brighter than or equal to the observed value (i.e., $M_{\mathrm{rep}} \leq M_{\mathrm{obs}}$, given the inverse logarithmic nature of the magnitude scale). The complementary probability of obtaining a fainter magnitude is $1 - F(M_{\mathrm{obs}})$. In practice, $F(M_{\mathrm{obs}})$ is estimated empirically as the fraction of simulated replicated magnitudes satisfying $M_{\mathrm{rep}} \leq M_{\mathrm{obs}}$.

The two-sided tail-area probability is then defined as:

\begin{equation}
P_{\mathrm{tail},\mathrm{KNe}} = 2 \min \big( F(M_{\mathrm{obs}}), 1 - F(M_{\mathrm{obs}}) \big) \, .
\end{equation}

Therefore, $P_{\mathrm{tail},\mathrm{KNe}} \in [0,1]$, where smaller values indicate that the observation lies in the extreme tails of the PPD (low plausibility), and values closer to unity indicate that the observation is typical of the model’s predictions (high plausibility).

To quantify the uncertainty associated with the $P_{\mathrm{tail},\mathrm{KNe}}$ score, we propagate the observational uncertainty by sampling the observed magnitude distribution. Specifically, we generate $N_{\rm obs} = 100$ realizations of the observed absolute magnitude, drawing each from a Gaussian distribution centered at $M_{\mathrm{obs}}$ with standard deviation $\sigma_{\mathrm{obs}}$. For each realization, we compute the corresponding $P_{\mathrm{tail},\mathrm{KNe}}$ value. This procedure yields an empirical distribution of $P_{\mathrm{tail},\mathrm{KNe}}$ scores, enabling us to quantify its uncertainty.

Each photometric measurement is initially assigned an independent $P_{\mathrm{tail},\mathrm{KNe}}$ score along with its associated uncertainty $\sigma_{\mathrm{tail},\mathrm{KNe}}$. As new observations are obtained on subsequent nights, potentially supplemented by additional follow-up data, or when multiple observations are available within the same temporal bin of width $0.2$~MJD, it is necessary to combine the individual scores into a cumulative score. 

To achieve this, we adopt a sequential update scheme that first aggregates scores within the same time bin using an inverse-variance weighted mean in logit space. This aggregation accounts for the uncertainty associated with each individual score and yields a more stable and reliable estimate of model consistency at that epoch. Because the consistency scores are bounded in $[0,1]$ and can exhibit large relative uncertainties near the boundaries, performing the aggregation in logit space stabilizes variances and prevents extreme scores with small uncertainties from disproportionately influencing the cumulative score. 

For instance, consider two observations on the same night with individual $P_{\mathrm{tail},\mathrm{KNe}} = [0.8, 0.01]$ and uncertainties $\sigma = [0.1, 0.05]$. A straightforward inverse-variance weighted mean in probability space would yield a cumulative score of $0.168 \pm 0.044$, heavily biased toward the smaller value with lower \textit{absolute} uncertainty. In contrast, computing the inverse-variance weighted mean in logit space produces a cumulative score of $0.785 \pm 0.104$, correctly reflecting the dominant contribution of the higher score with the lower \textit{relative} uncertainty. 

This procedure is applied both for combining multiple observations within the same time bin and for sequentially updating the cumulative score as new observations become available. By operating in logit space, we obtain a more conservative and statistically robust aggregation of scores, which is particularly important during early observations when the number of detections is limited. This approach prevents a candidate from being prematurely excluded or ranked as dissimilar to a kilonova based on a single low score with a large relative uncertainty.

\textbf{The sequential update scheme:} The overall method can be summarized as follows. Let $x_i$ denote the $P_{\mathrm{tail},\mathrm{KNe}}$ score derived from the $i$-th observation, with associated uncertainty $\sigma_i$. The inverse-variance weighted mean score in the logit space is formulated as:

\begin{equation}
z_i = \log\!\left(\frac{x_i}{1 - x_i}\right),
\end{equation}
with the corresponding uncertainty propagated using the delta method,
\begin{equation}
\sigma_{z_i} \approx \frac{\sigma_i}{x_i(1-x_i)}.
\end{equation}

The inverse-variance weighted mean in logit space is then defined as
\begin{equation}
\bar{z}_{\mathrm{w}} =
\frac{\displaystyle \sum_{i=1}^{N} \frac{z_i}{\sigma_{z_i}^{2}}}
     {\displaystyle \sum_{i=1}^{N} \frac{1}{\sigma_{z_i}^{2}}},
\end{equation}
with variance
\begin{equation}
\sigma_{\bar{z}_{\mathrm{w}}}^{2}
=
\left(
\sum_{i=1}^{N} \frac{1}{\sigma_{z_i}^{2}}
\right)^{-1}.
\end{equation}

The aggregated score in the original domain is obtained via the inverse logit (sigmoid) transform,
\begin{equation}
\bar{x}_{\mathrm{w}} = \frac{1}{1 + e^{-\bar{z}_{\mathrm{w}}}}.
\end{equation}

This formulation admits a sequential update scheme in which a previous cumulative estimate $(\bar{z}_{\mathrm{prev}}, \sigma_{\mathrm{prev}})$ is updated with a new observation $(z_{\mathrm{new}}, \sigma_{\mathrm{new}})$ according to
\begin{align}
\bar{z}_{\mathrm{new}} &=
\frac{
\bar{z}_{\mathrm{prev}}/\sigma_{\mathrm{prev}}^{2}
+
z_{\mathrm{new}}/\sigma_{\mathrm{new}}^{2}
}{
1/\sigma_{\mathrm{prev}}^{2}
+
1/\sigma_{\mathrm{new}}^{2}
}, \\[6pt]
\sigma_{\bar{z}_{\mathrm{new}}}^{2} &=
\left(
\frac{1}{\sigma_{\mathrm{prev}}^{2}}
+
\frac{1}{\sigma_{\mathrm{new}}^{2}}
\right)^{-1},
\end{align}
where $\sigma_{\bar{z}_{\mathrm{new}}}^{2}$ is the variance of the inverse-variance pooled logit estimate $\bar{z}_{\mathrm{new}}$, distinct from the per-observation score uncertainty $\sigma_{\mathrm{new}}^{2}$.

This sequential aggregation enables real-time updating of the cumulative kilonova consistency score as new observations are acquired. In the context of searching for rare and rapidly evolving transients such as kilonovae, it is preferable to tolerate a higher rate of false positives at early stages than to risk rejecting good candidates due to overly restrictive scoring criteria. The scoring framework is therefore designed to retain plausible kilonova candidates while producing a more robust ranking at later stages with additional data.

\subsubsection{\texorpdfstring{$P_{\mathrm{near},\mathrm{KNe}}$ Score}{P-near KNe Score}} \label{sec:pnear}

As a complementary measure of model consistency, we define the $P_{\mathrm{near},\mathrm{KNe}}$ 
score, which quantifies the local agreement between the observed magnitude and the 
model-predicted distribution. This score is inspired by the Region of Practical Equivalence 
(ROPE, \citealt{rope_Kruschke2013, rope_Kruschke2018}) framework, commonly used in equivalence testing. The ROPE is defined as a symmetric interval around a 
reference value within which differences are deemed practically negligible. Formally, for a 
parameter $\theta$ and a tolerance $\delta$, $\mathrm{ROPE}(\theta_0) = [\theta_0 - \delta,\, 
\theta_0 + \delta]$, and a parameter estimate is considered practically equivalent to $\theta_0$ 
if its posterior mass within the ROPE exceeds a prescribed threshold \citep{rope_Kruschke2018}. 
ROPE is typically applied in parameter space to declare practical equivalence. Here we 
adapt it to observable space as a tolerance region for prior predictive checking, centering the interval on the observed magnitude $M_{\mathrm{obs}}$ rather than on a reference parameter value.

The $P_{\mathrm{near},\mathrm{KNe}}$ score is defined as the probability mass of the 
noise-convolved prior predictive distribution of replicated magnitudes, $M_{\mathrm{rep}}$, 
that falls within a symmetric tolerance interval centered on the observed magnitude 
$M_{\mathrm{obs}}$. Formally, we define the ROPE in observable space as
\begin{equation}
    \mathcal{R}(M_{\mathrm{obs}}, \sigma_{\mathrm{obs}}) =
    \left[
    M_{\mathrm{obs}} - k_{\mathrm{near}}\,\sigma_{\mathrm{obs}},\;
    M_{\mathrm{obs}} + k_{\mathrm{near}}\,\sigma_{\mathrm{obs}}
    \right],
\end{equation}

where $\sigma_{\mathrm{obs}}$ denotes the propagated observational uncertainty and $k_{\mathrm{near}}$ is a 
user-defined tolerance factor governing the width of the equivalence region.\footnote{In this work we adopt $k_{\mathrm{near}} = 3.0$ and $k_{\mathrm{ABC}} = 1.5$ as fiducial values.}

The $P_{\mathrm{near},\mathrm{KNe}}$ score is then defined as

\begin{equation}
\begin{split}
    P_{\mathrm{near},\mathrm{KNe}} &= \Pr\!\left( M_{\mathrm{rep}} \in \mathcal{R}(M_{\mathrm{obs}}, \sigma_{\mathrm{obs}}) \right) \\
    &= \int_{\mathcal{R}} p(M_{\mathrm{rep}} \mid t, b)\, \mathrm{d}M_{\mathrm{rep}},
\end{split}
\end{equation}

where $p(M_{\mathrm{rep}} \mid t,b)$ is the prior predictive distribution of 
noise-convolved magnitudes for a given time and band. The score thus measures the local density of the PPD in the neighborhood of the observation. In practice, $P_{\mathrm{near},\mathrm{KNe}}$ is estimated empirically from a finite set of 
$N_{\mathrm{samples}}$ Monte Carlo draws $\{M_{\mathrm{rep},i}\}_{i=1}^{N_{\mathrm{samples}}}$ from the PPD as

\begin{equation}
    \widehat{P}_{\mathrm{near},\mathrm{KNe}} =
    \frac{1}{N_{\mathrm{samples}}} \sum_{i=1}^{N_{\mathrm{samples}}}
    \mathbf{1}\!\left[ M_{\mathrm{rep},i} \in \mathcal{R}(M_{\mathrm{obs}},
    \sigma_{\mathrm{obs}}) \right],
\end{equation}

where $\mathbf{1}[\cdot]$ is the indicator function. Thus, $\widehat{P}_{\mathrm{near},\mathrm{KNe}}$ is the sample fraction of replicated draws falling within the ROPE. High values of $P_{\mathrm{near},\mathrm{KNe}}$ indicate that the PPD places substantial probability mass near $M_{\mathrm{obs}}$, reflecting strong local agreement between the observation and the model. Conversely, low values signal that the observation falls in a 
region of low density, such as a tail or a local minimum of the PPD.

Unlike the $P_{\mathrm{tail},\mathrm{KNe}}$ score, no additional uncertainty is associated 
with $P_{\mathrm{near},\mathrm{KNe}}$: the observational uncertainty $\sigma_{\mathrm{obs}}$ 
is already encoded in two ways: \emph{first}, in the ROPE width $\mathcal{R}$, which defines the integration limits of the score, and \emph{second}, in the noise convolution of the PPD, which accounts for photometric measurement error. Furthermore, $P_{\mathrm{near},\mathrm{KNe}}$ is evaluated independently for each 
observation and is not aggregated across bands or epochs, reflecting its role as a \emph{local} consistency measure rather than a cumulative score \citep{rope_Gelman2013}.

\subsection{Approximate Bayesian Computation Diagnostic} \label{sec:abc_diagnostic}

Approximate Bayesian Computation (ABC) is a class of likelihood-free inference methods 
designed to estimate posterior distributions when the likelihood function is computationally 
intractable or analytically unavailable \citep{abc_Sisson2007,abc_DelMoral2011,Marin2012ABCreview}. Rather than evaluating the likelihood directly, ABC proceeds by simulating light curves from the kilonova model for different parameter values and retaining only those simulations whose distance from the observed data falls below a predefined threshold $\varepsilon$, the \emph{acceptance kernel}. 

In this work, we adopt the principles of ABC to construct a \emph{diagnostic} tool for 
the temporal consistency of kilonova candidates, rather than to perform full Bayesian 
inference. The ROPE defined in Section~\ref{sec:pnear} acts as the acceptance kernel, with 
the discrepancy measure being the absolute difference between the simulated and 
observed magnitude at each epoch and band. A simulation is retained at epoch $t$ if its predicted 
magnitude falls within $\mathcal{R}(M_{\mathrm{obs}}, k_{\mathrm{ABC}}\times\sigma_{\mathrm{obs}})$. Otherwise it is rejected.

\textbf{Sequential Survival Filtering:} Rather than treating each epoch as an independent event, we implement a sequential survival diagnostic: beginning from the first observation, we identify the subset $\mathcal{S}_1$ of simulations consistent with the initial ROPE constraint. As each subsequent observation is incorporated across all available photometric bands, only 
simulations that satisfy \emph{all} previous ROPE criteria are retained. Formally, the 
surviving set at epoch $t$ is
\begin{equation}
    \mathcal{S}_t = \mathcal{S}_{t-1} \cap
    \left\{
        i : M_{\mathrm{rep},i}(t) \in \mathcal{R}(M_{\mathrm{obs}}(t),
        \sigma_{\mathrm{obs}}(t))
    \right\},
\end{equation}
with $|\mathcal{S}_t| \leq |\mathcal{S}_{t-1}|$ by construction, so that the survival 
count is monotonically non-increasing. This sequential structure is analogous to an ABC sequential Monte Carlo (SMC) scheme \citep{abc_DelMoral2011,abc_Sisson2007}, in which the posterior is refined iteratively as additional data are incorporated. However, unlike standard ABC-SMC, which refines the 
posterior by progressively tightening the acceptance threshold $\varepsilon$, here the 
ROPE width remains fixed at $k_{\mathrm{ABC}}=1.5$, this choice is discussed on Appendix \ref{app:settings}. This ensures a conservative diagnostic tailored to detect temporal inconsistency in the observed light curve relative to the model grid, rather than to perform parameter estimation.

The relative survival fraction at epoch $t$ is defined as the number of 
simulations surviving all cumulative constraints up to epoch $t$, normalized 
by the number of simulations independently consistent with the observation 
at that epoch alone,
\begin{equation}
    f_{\mathrm{surv}}(t) = \frac{|\mathcal{S}_t|}{|\mathcal{A}_t|+\epsilon},
\end{equation}
where $|\mathcal{S}_t|$ is the size of the sequential surviving set (simulations satisfying the ROPE criterion at \emph{all} epochs up to and
including $t$), and $|\mathcal{A}_t|$ is the number of simulations 
independently accepted at epoch $t$ alone, without any prior 
constraints and $\epsilon$ is a small constant to ensure numerical stability. 
By construction, $|\mathcal{S}_t| \leq |\mathcal{A}_t|$, constraining $f_{\mathrm{surv}}(t) \in [0, 1]$. This fraction serves as a measure of physics-informed temporal coherence: a value of $f_{\mathrm{surv}}(t) \approx 1$ indicates that nearly all simulations consistent with the current observation also align with the entire photometric history. Conversely, $f_{\mathrm{surv}}(t) \rightarrow 0$ signals that while the current epoch in isolation is explained by simulations, almost none survive the full sequential filter, indicating a fundamental tension between the current observation and the preceding photometric evolution. 

The expected behavior of $f_{\mathrm{surv}}(t)$ depends on the nature of the transient 
and the information content of the observations. For a physically consistent kilonova, 
the survival fraction is expected to decrease monotonically as new data progressively 
constrain the viable model space, eventually reaching a plateau that reflects the 
irreducible degeneracy of the model grid given the available data. 

For a contaminant such as a supernova, which may mimic kilonova photometric evolution at 
early epochs but diverge significantly as the transient evolves, the survival fraction is 
expected to collapse to zero as the cumulative constraint becomes incompatible with any 
single model in the grid. A collapse $|\mathcal{S}_t| \to 0$ indicates a fundamental 
divergence: no simulation within the prior grid can simultaneously explain all observed 
epochs, even if individual observations appear kilonova-like in isolation. 

\textbf{Penalization and Flagging:} In cases where $|\mathcal{S}_t| = 0$ at any epoch $t$, the candidate is flagged as temporally inconsistent with the kilonova model grid. Even if a high $P_{\mathrm{near}, 
\mathrm{KNe}}$ or $P_{\mathrm{tail},\mathrm{KNe}}$ score is recorded for a subsequent 
isolated observation, the cumulative kilonova score is penalized and set to zero. This 
hard penalization reflects the logical requirement that a valid kilonova candidate must be 
globally consistent with at least one kilonova model across all epochs, and prevents spuriously high scores arising from chance alignment for an individual observation \citep{ppcGelman1996}.

Given our adoption of broad priors intended to span a wide range of kilonova-like transients, together with a choice of $k_{\mathrm{ABC}} = 1.5$ and a noise-convolved prior predictive distribution (PPD), the absence of surviving samples constitutes strong evidence that either the transient is inconsistent with a kilonova interpretation or that the simulated model grid does not adequately capture the observed photometric evolution.

\section{Results \& Validation}\label{sec:results}

We validate \pkg{KilonovaSCORER} against four independent test cases of increasing complexity. First, we score AT\,2017gfo, the electromagnetic counterpart of GW170817, which remains the only confirmed kilonova detected in direct association with a gravitational-wave event. Second, we study SN\,2025ulz, a Type IIb supernova that evolved similarly to a kilonova during the first few days after the trigger and was initially misidentified as a candidate counterpart to S250818k. Third, we apply the pipeline to published light curves of kilonovae associated with GRBs, which by selection are observed at small inclination angles relative to the jet axis and high redshifts. These provide an additional consistency check under viewing geometries distinct from that of GW170817. Fourth, as a statistical validation, we score a simulated population of transients (including kilonovae from both BNS and NSBH progenitors, Type Ia supernovae, and Type II supernovae) generated under the LSST limiting magnitude and the silver ToO observing strategy.

\textbf{Candidate Diagnostic Report:} Throughout the remainder of this section, results are presented using a standardized Candidate Diagnostic Report designed to facilitate rapid visual vetting of transient candidates. Each report consists of four panels. The \textit{top panel (a)} shows the candidate photometry in absolute magnitude as a function of time since merger, overlaid on the surviving population of kilonova simulations color-coded by filter band. Individual detections are rendered as band-specific markers colored by their $P_{\mathrm{near},\mathrm{KNe}}$ score.
The \textit{second panel (b)} displays the $P_{\mathrm{near},\mathrm{KNe}}$ score for each individual detection as a function of time. The dashed horizontal line marks an empirical reference below which fewer than 20\% of the simulations fall within the observation's ROPE. 
The \textit{third panel (c)} shows the relative survival fraction $f_{\mathrm{surv}}(t)=|\mathcal{S}_t|/(|\mathcal{A}_t|+\epsilon)$ from the ABC diagnostic (Section~\ref{sec:abc_diagnostic}) on a logarithmic scale as a function of time since merger, with each point colored by the number of accepted simulations ($|\mathcal{A}_t|$) at that epoch based on the acceptance criteria from the ABC diagnostic. The presence of a vertical red line signals when $|\mathcal{S}_t| \to 0$. 
The \textit{bottom panel (d)} presents the cumulative score evolution ($P_{\mathrm{tail},\mathrm{KNe}}$), showing the per-epoch bin score (light gray dot) alongside the running cumulative score (purple square) with its uncertainty as shaded region. The number of detections contributing to each per-epoch bin score is annotated directly above the corresponding point.

\subsection{AT\,2017gfo \& GW170817}

We begin by applying \pkg{KilonovaSCORER} to AT\,2017gfo as a benchmark
validation, given that it remains the only confirmed electromagnetic counterpart
to a binary neutron star merger. We expect our framework to assign consistently
high scores to this event, particularly during the first $\sim$5 days, especially before the
peak brightness, when rapid candidate prioritization is most critical to enable more informative multi-wavelength and spectroscopic follow-ups. 

\begin{figure}[!htbp]
    \centering
    \includegraphics[width=\linewidth]{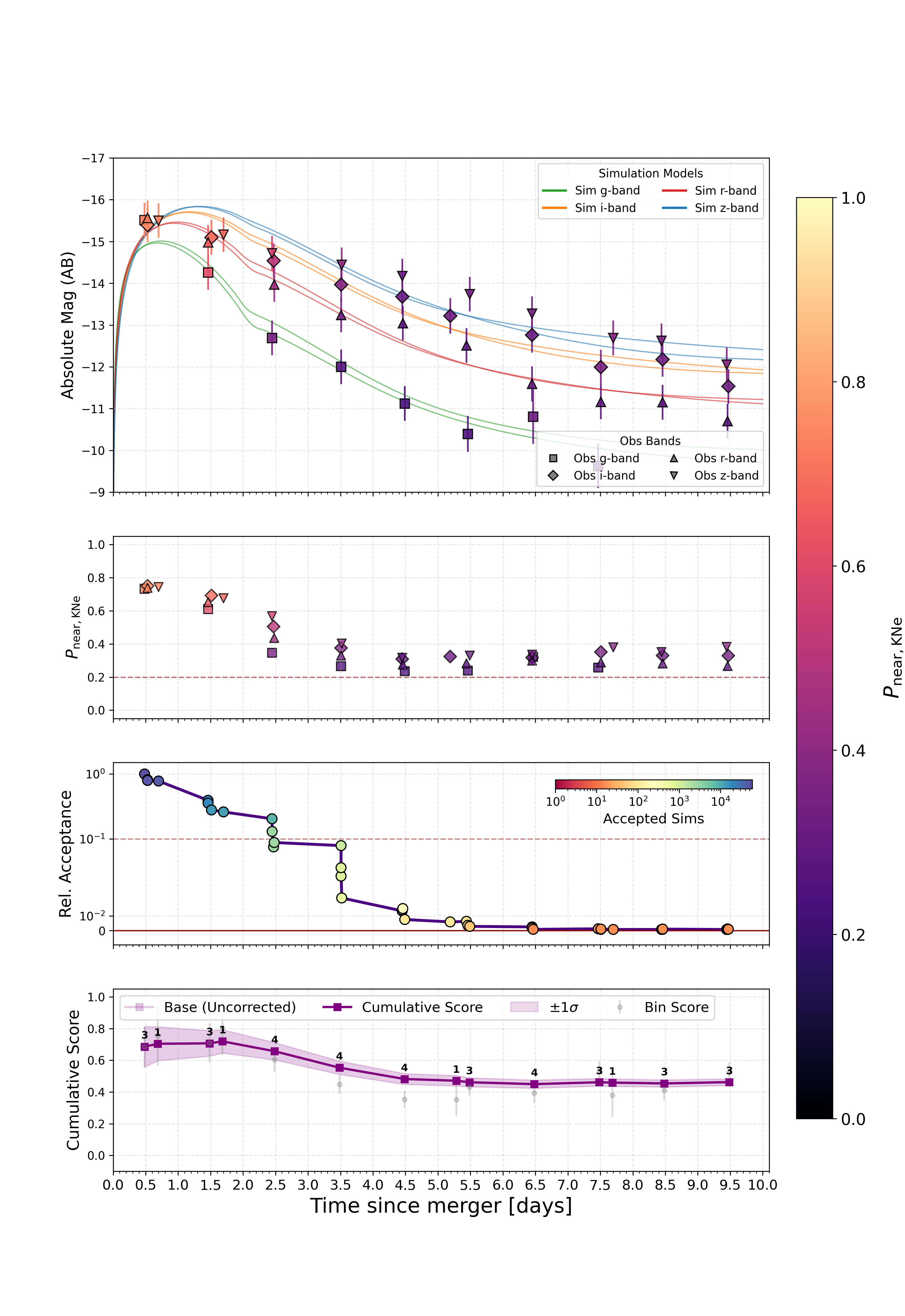}
    \caption{Candidate Diagnostic Report for AT\,2017gfo scored in the $griz$ 
    bands. \textit{Top panel:} Multi-band light curve in absolute magnitude overlaid on the surviving 
    population of kilonova simulations, with individual detections color-coded 
    by $P_{\mathrm{near},\mathrm{KNe}}$ score. \textit{Second panel:} Per-epoch 
    $P_{\mathrm{near},\mathrm{KNe}}$ score as a function of time since merger. 
    \textit{Third panel:} Relative survival fraction $f_{\mathrm{surv}}(t)$ from the ABC diagnostic 
    on a logarithmic scale, with each point colored by the number of surviving 
    simulations. \textit{Bottom panel:} Cumulative score evolution showing the 
    per-epoch bin $P_{\mathrm{tail},\mathrm{KNe}}$ score (grey markers) 
    and the running cumulative score with its uncertainty.
    Time since merger on horizontal axes in all panels.}
    \label{fig:at2017gfo_all}
\end{figure}

\begin{figure}[!htbp]
    \centering
    \includegraphics[width=\linewidth]{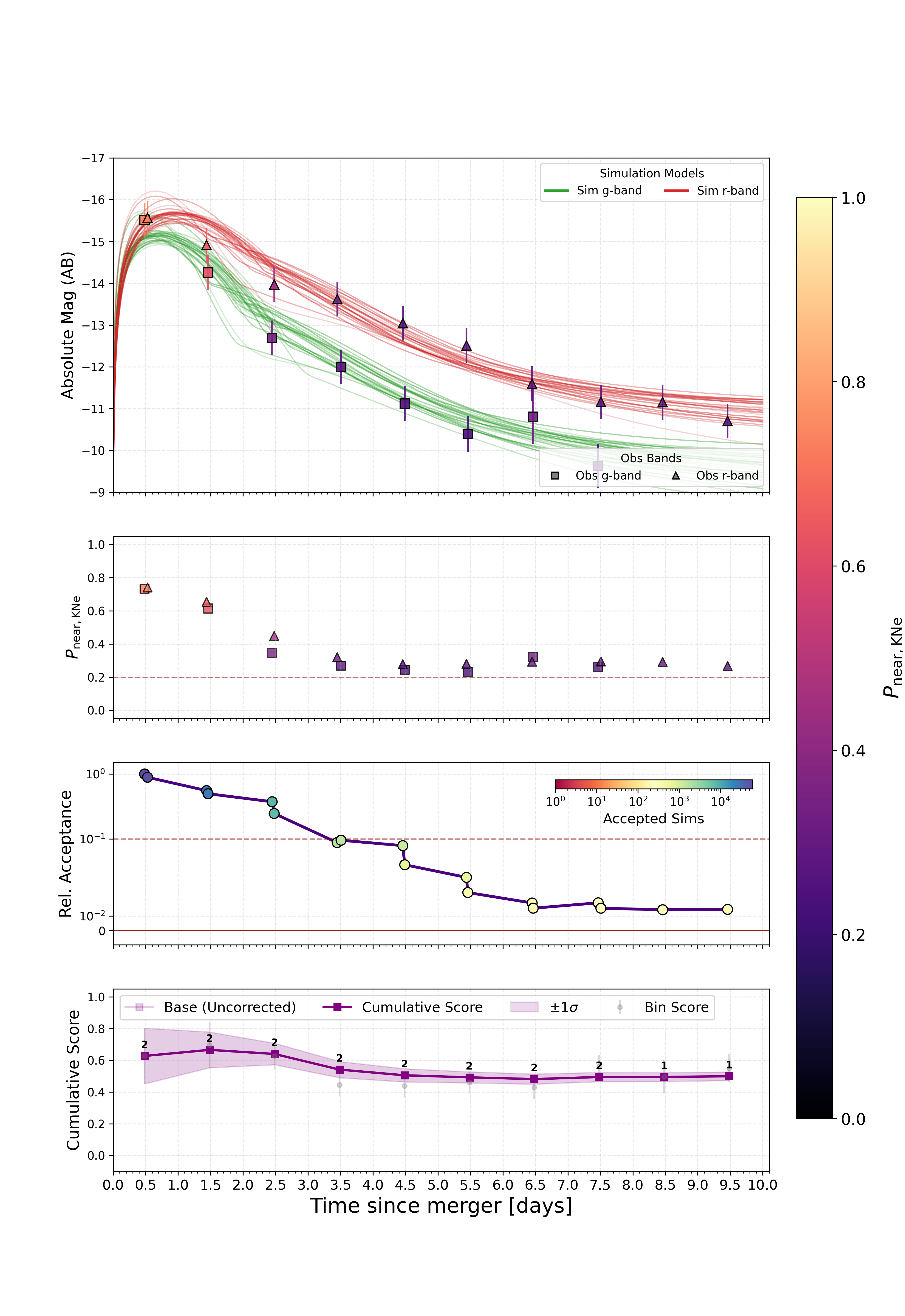}
    \caption{Same as Figure~\ref{fig:at2017gfo_all} but for AT\,2017gfo scored 
    under a two-band ($g+r$) silver ToO observing strategy.}
    \label{fig:at2017gfo_gr}
\end{figure}

To emulate a realistic EM counterpart search scenario, we analyze AT\,2017gfo as if it were an unknown candidate lying within the gravitational-wave 
localization volume, with no prior host galaxy association. Our analysis is 
restricted to LSST-like photometric bands ($g$, $r$, $i$, $z$) and to 
observations obtained within the first 10 days post-merger. Because AT\,2017gfo 
was exceptionally well-observed, its photometric cadence substantially exceeds 
what is expected for a typical candidate in a real EM counterpart search. We 
therefore downsample the dataset by retaining a single epoch per night per band, 
selecting the detection with the highest signal-to-noise ratio. We score two 
versions of this dataset: the full $griz$ coverage (Figure~\ref{fig:at2017gfo_all}) and a $g+r$ only version emulating the LSST silver ToO strategy, in which coverage is restricted to two observations per night 
(Figure~\ref{fig:at2017gfo_gr}).

In both Candidate Diagnostic Reports, panel (a) shows the multi-band light curve
of AT\,2017gfo overlaid on the subset of simulated kilonova light curves retained
by the ABC sequential filter (Section~\ref{sec:abc_diagnostic}).
\pkg{KilonovaSCORER} performs as expected: at early times ($t \lesssim 2$
days), all detections received a $P_{\mathrm{near},\mathrm{KNe}} > 0.6$, indicating
that more than 60\% of the simulated light curves fall within the region of
practical equivalence $\mathcal{R}(M_\mathrm{obs}, \sigma_\mathrm{obs})$ of the
observed photometry (panel b of Figures~\ref{fig:at2017gfo_all}
and~\ref{fig:at2017gfo_gr}). The cumulative score evolves stably across the
first three nights in both scenarios. For the full $griz$ coverage, the score was $\sim$$0.6 \pm 0.1$ in the first night, to $\sim$$0.55
\pm 0.08$ after night two, and $\sim$$0.54 \pm 0.08$ after night three, 
precisely the window in which candidate ranking is most valuable and
spectroscopic follow-up resources are most urgently needed. As subsequent
observations are taken into account, the cumulative
score stabilizes at $0.44 \pm 0.05$. Correspondingly, the relative survival
fraction ($f_\mathrm{surv}(t) = |\mathcal{S}_t|/(|\mathcal{A}_t|+\epsilon)$) declines
monotonically but never reaches zero (panel c of Figures~\ref{fig:at2017gfo_all}
and~\ref{fig:at2017gfo_gr}), the behavior expected of a genuine kilonova.

\subsection{SN\,2025ulz \& S250818k}

SN\,2025ulz was detected and ingested into the \pkg{TROVE} system as a candidate
counterpart to the gravitational-wave event S250818k. Its early photometric
evolution closely resembled that of a kilonova, motivating its initial
classification as a potential kilonova and subsequent studies proposing it as
an unusually luminous ``superkilonova'' \citep{ulz_Kasliwal2025}. Subsequent
spectroscopic observations established that SN\,2025ulz is more consistently
explained as a Type~IIb supernova \citep{Franz2025}. This source therefore
constitutes an ideal validation case: a transient that genuinely mimics
kilonova-like photometric behavior at early times but diverges at later epochs,
allowing us to assess whether \pkg{KilonovaSCORER} correctly down-ranks it
once the divergence becomes apparent in the data.

\begin{figure}[!htbp]
    \centering
    \includegraphics[width=\linewidth]{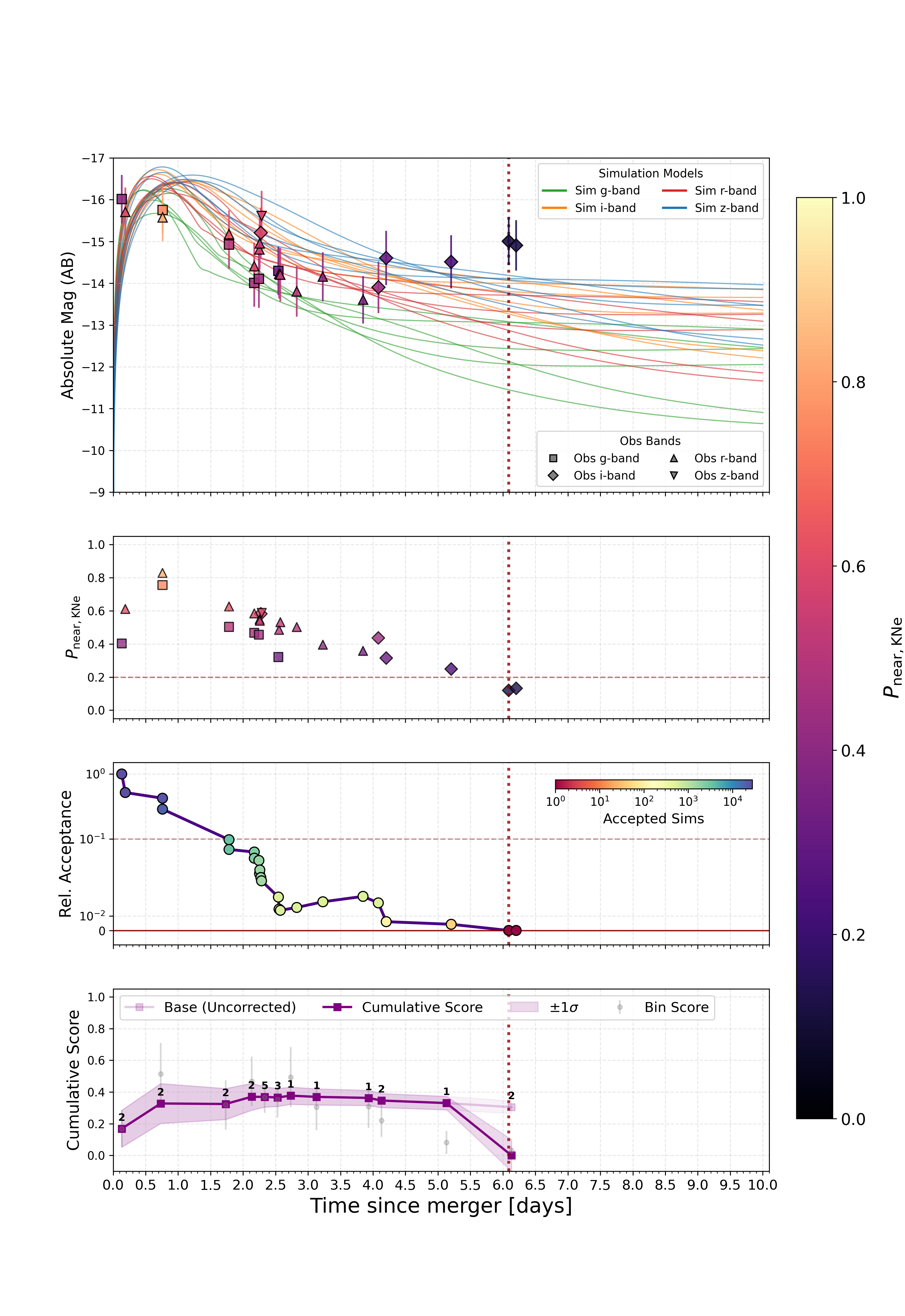}
    \caption{Candidate Diagnostic Report for SN\,2025ulz, a Type~IIb supernova
    identified as a candidate counterpart to S250818k. See
    Figure~\ref{fig:at2017gfo_all} for a description of each panel. The
    vertical dashed line in panel (c) marks the epoch at which the survival
    fraction reaches zero, after which the cumulative score is hard-penalized
    to zero.}
    \label{fig:at2025ulz}
\end{figure}

Figure~\ref{fig:at2025ulz}(a) shows the multi-band light curve of SN\,2025ulz
overlaid on the subset of simulated kilonova light curves retained by the ABC
sequential filter before the collapse time (vertical red-dashed line) (Section~\ref{sec:abc_diagnostic}). During the first
${\sim}4$ days, the photometric evolution is broadly consistent with the
kilonova model grid, and the surviving simulation set remains non-empty
(panel c), consistent with the early misclassification of this event as a
kilonova candidate. At $t \sim 4$--$6$ days, however, a clear re-brightening
emerges in the $i$-band that is incompatible with the monotonically declining
post-peak evolution expected for a kilonova. Although the detections associated
with the re-brightening individually yield non-zero $P_{\mathrm{near},
\mathrm{KNe}}$ and $P_{\mathrm{tail},\mathrm{KNe}}$ scores, indicating that
the observed magnitudes are not implausible at those epochs in isolation, 
these per-epoch scores are misleading, as they do not encode the temporal
coherence of the light curve. This is precisely the temporal behavior that the ABC diagnostic is designed to flag.

The ABC diagnostic captures this inconsistency explicitly through its
sequential structure: the relative survival fraction $f_{\mathrm{surv}}(t) =
|\mathcal{S}_t|/(|\mathcal{A}_t|+\epsilon)$ declines sharply once the first observation that indicates a re-brightening is incorporated into the dataset and reaches $|\mathcal{S}_t| = 0$ at $t \sim
6$ days, at which point no simulated kilonova remains consistent with the
i-band observed magnitude of $M_{\mathrm{obs}} = -14.89 \pm 0.58$\,mag (vertical
dashed line, panel c). SN\,2025ulz is thereby flagged as temporally inconsistent
with the kilonova model grid, and the cumulative score is hard-penalized to
zero for all subsequent epochs (Section~\ref{sec:abc_diagnostic}). 

The cumulative score evolution (panel d) further reflects the ambiguous nature
of this transient. During the first ${\sim}4$ days, as observations from
multiple follow-up teams are ingested nightly into \pkg{TROVE}, individual epochs
with $P_{\mathrm{tail},\mathrm{KNe}} > 0.4$ drive the cumulative score
upward, reaching a peak of $0.38 \pm 0.05$ before the re-brightening is
detected. This intermediate score is itself informative: it is non-negligible,
reflecting the genuine photometric similarity to a kilonova at early times, yet
it remains below the scores achieved by AT\,2017gfo over the same window. Once $|\mathcal{S}_t| \to 0$ is registered, the
cumulative score is hard-penalized to zero, correctly identifying and ranking SN\,2025ulz as
inconsistent with a kilonova and demonstrating that \pkg{KilonovaSCORER}
can discriminate kilonova impostors even when per-epoch scores remain
superficially plausible.

\subsection{Kilonovae Associated with Gamma-Ray Bursts}

\begin{figure}[!htbp]
    \centering
    \includegraphics[width=0.99\linewidth]{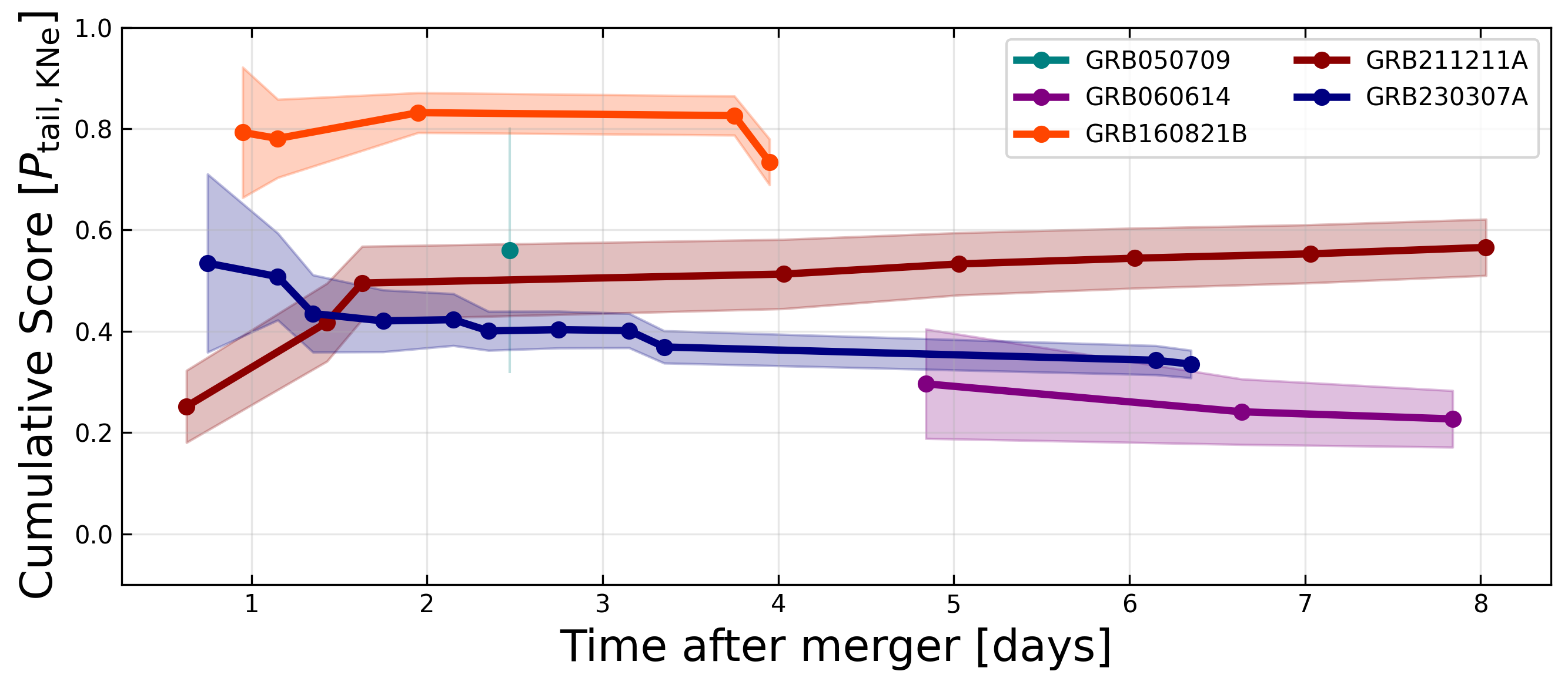}
    \caption{Cumulative $P_{\mathrm{tail},\mathrm{KNe}}$ scores as a function 
    of time since merger (days) for five kilonovae associated with GRB events. The scores combine the optical ($g$,$r$,$i$,$z$) and IR ($K$,$H$) bands. The vertical axis is the running cumulative score in $[0,1]$.}
    \label{fig:grb}
\end{figure}

Alongside the GW search, we validate \pkg{KilonovaSCORER} against 
a sample of confirmed kilonovae associated with gamma-ray 
bursts. These events offer a valuable independent test: they are confirmed 
kilonovae at known distances, observed under viewing geometries and luminosity scales largely different from AT\,2017gfo, thereby probing a distinct region of the kilonova parameter space sampled by our simulation grid. The observed optical light curves of GRB counterparts are generally composed of multiple emission components, including an afterglow component and a kilonova contribution that may itself consist of multiple ejecta components. In recent work, the afterglow emission has been constrained independently using gamma-ray and radio observations, allowing it to be modeled and subtracted from the optical light curve \citep{Rastinejad2025}. This procedure isolates the kilonova component and makes the residual light curves directly comparable to our simulation grid.

We compiled afterglow-subtracted optical photometry for seven GRB-associated kilonova events from \citet{Rastinejad2025}, summarized in 
Table~\ref{tab:grb_sample}. Two are associated with long-duration GRBs: 
GRB\,230307A \citep{LevanGRB230307A,Yang2024NatureKNeLGRB} and GRB\,211211A 
\citep{RastinajGRB211211A}, and the remaining five are associated with short GRBs. 

GRB-selected kilonovae are typically observed at small inclination 
angles relative to the jet axis, allowing the optical/IR counterpart to be observable at luminosity distances (up to $\sim 3300\,\mathrm{Mpc}$) greater than those generally accessible in GW-based searches. This selection effect naturally biases the sample toward intrinsically more luminous systems. 

Our comparison implicitly assumes equivalence between observed and rest-frame photometric bands. While this approximation is generally adequate for GW follow-up observations, it becomes increasingly inaccurate for GRB-selected events at higher redshift. In this regime, the shift in effective rest-frame wavelength, $\lambda_{\mathrm{rest}} = \lambda_{\mathrm{obs}}/(1+z)$, leads to non-negligible band mismatches, and proper K-corrections, rest-frame filter mapping, and time-dilation corrections should be applied. For example, the r-band ($\lambda \sim 6200\,\AA$) at $z \approx 0.3$ maps to $\sim 4770\,\AA$, closer to the g-band, with similar effects for i- and z-bands at $z \gtrsim 0.2$. GRB130603B and GRB200522A would require band mapping. However, both events have only a few post-peak IR (F150W and F115W) observations from the Hubble Space Telescope, outside the temporal regime relevant for testing the early-time ranking performance of \pkg{KilonovaSCORER}. We therefore exclude these events from our analysis.

GRB160821B lies near the redshift regime where band remapping between observer-frame and rest-frame filters becomes relevant. To quantify this effect, we computed a cross-correlation matrix between filter groups by approximating each bandpass as a top-hat (rectangle) function in wavelength and evaluating the fractional overlap between rest-frame filters and their redshifted observer-frame counterparts. We find that several filters, particularly at longer wavelengths, require a systematic shift toward bluer rest-frame bands. However, when incorporating this mapping into the scoring procedure, the resulting changes in the individual scores are at order of $\sim$0.1--0.2 \citep{2021ApJ_grb_rastinejad,rastinejad2024uniformmodelingobservedkilonovae}. In practice, the cumulative score is slightly higher when no explicit band remapping is applied. This behavior indicates that \pkg{KilonovaSCORER}, by construction, is relatively insensitive to small mismatches between observer-frame and rest-frame bandpasses at redshifts $\lesssim 0.16$. Its conservative scoring scheme therefore remains robust for GRB-associated transients, without requiring explicit redshift corrections, including filter transformations, K-corrections, or time dilation effects, for the current sample.

\begin{table}[t]
\centering
\caption{Summary of kilonovae associated with GRBs used in this work,
compiled from \citet{rastinejad2024uniformmodelingobservedkilonovae}.
Listed are the GRB identifier, redshift ($z$), and luminosity distance
($D_{\mathrm{L}}$). An asterisk (*) denotes kilonovae associated with
long-duration GRBs.}
\label{tab:grb_sample}
\begin{tabular}{lcc}
\hline\hline
GRB ID & $z$ & $D_{\mathrm{L}}\;(\mathrm{Mpc})$ \\
\hline
050709   & 0.1610 &  796.31 \\
060614   & 0.1250 &  604.53 \\
130603B  & 0.3560 & 1954.14 \\
160821B  & 0.1616 &  799.57 \\
200522A  & 0.5536 & 3299.25 \\
211211A$^{*}$ & 0.0763 &  357.30 \\
230307A$^{*}$ & 0.0650 &  302.02 \\
\hline
\end{tabular}
\end{table}

In the absence of a gravitational-wave detection, no direct distance measurement is available, and we therefore adopt the host galaxy redshift as a distance proxy, assuming a 10\% distance uncertainty comparable to typical GW-derived estimates, using a Planck~2015 $\Lambda$CDM cosmology \citep{Planck2016}. 

For the remaining five GRB events with associated kilonovae, we extend \pkg{KilonovaSCORER} to operate across optical and infrared (IR) bands ($g$, $r$, $i$, $z$, $J$, $H$, $K$, F356W, F444W), mapping each observed bandpass to the nearest corresponding band group (e.g., F160W $\rightarrow$ $H$-band, F115W $\rightarrow$ $J$-band). Figure~\ref{fig:grb} shows the cumulative score evolution for all five GRB-associated kilonovae. \pkg{KilonovaSCORER} assigns a cumulative score greater than 0.2 to every event in the sample; the short GRBs GRB\,160821B and GRB\,050709 achieve the highest cumulative scores of $0.83 \pm 0.07$ and $0.56 \pm 0.24$, respectively, near the emission peak.

The cumulative score evolution of GRB\,211211A reflects that the earliest observations are substantially more luminous than the bulk of the kilonova simulation grid, and seven detections fall within the same temporal bin, together yielding a cumulative score of $0.25 \pm 0.07$ at early times. Despite this, detections on the following night produce individual $P_{\mathrm{tail},\mathrm{KNe}}$ scores above 0.8, and subsequent $J$- and $K$-band photometry drives the cumulative score upward as the light curve evolves into better agreement with the expected kilonova population. Diagnostic reports for all five GRB events are provided in Appendix~\ref{sec:grb_appendix}. Across the full sample, all events consistently maintain individual $P_{\mathrm{near},\mathrm{KNe}}$ scores exceeding 0.1 throughout their observed evolution. For events with pre-peak detections, the cumulative $P_{\mathrm{tail},\mathrm{KNe}}$ enables these candidates to be ranked as high priority within two days of the trigger, with the cumulative score surpassing 0.4.

We note that while the mean afterglow contribution has been subtracted by \citet{Rastinejad2025}, residual afterglow contamination cannot be excluded entirely and may contribute to the elevated early-time fluxes observed in some events. Nevertheless, \pkg{KilonovaSCORER} does not reject any candidate in this sample: no event receives a zero individual or cumulative score, and the ABC diagnostic does not collapse within the first four days for any of the five events. These results demonstrate that \pkg{KilonovaSCORER} is not limited to scoring only optical bandpasses anticipated for LSST follow-up observations but can be naturally extended to IR wavelengths. This flexibility makes the ranking framework directly applicable to the Nancy Grace Roman Space Telescope \citep{ROMAN_KNe_ANDREONI2024102904}, whose wide-field NIR imaging capability uniquely complements ground-based searches.  In particular, Roman is well suited for discovering kilonovae at larger distances, with higher lanthanide fractions, larger binary mass ratios, or significant line-of-sight dust extinction. Applying \pkg{KilonovaSCORER} to Roman photometry would therefore provide a physics-informed ranking framework capable of operating seamlessly with Roman's observing strategy.

\subsection{Simulated Transient Populations: Supernova \& Kilonova}\label{sec:othertransients}

To evaluate the performance of \pkg{KilonovaSCORER} under realistic ToO conditions, we simulate the photometric evolution of the most common contaminants expected during gravitational-wave follow-up campaigns: core-collapse supernovae (CCSNe) and thermonuclear (Type Ia) supernovae. We additionally simulate kilonova light curves from
BNS and NSBH mergers as the reference class.

Simulated light curves for both supernova contaminants and kilonova reference signals are generated using \pkg{redback} \citep{REDBACKSarin2024}, with all model priors set to their default values (see \pkg{redback} documentation for details). Three supernova classes are considered, each modeled by a distinct physical mechanism and energy source: thermonuclear supernovae (Type Ia) are modeled using the \pkg{Arnett} model \citep{Arnett1982}, in which the light curve is powered entirely by $^{56}$Ni\,$\rightarrow$\,$^{56}$Co radioactive decay. Hydrogen-rich or partially stripped core-collapse supernovae (Types II, IIb, and some Ib/c) are modeled using \pkg{Arnett\_shock\_cooling} \citep{Piro2021_shock_cooling}, which includes an early emission component of shock-heated material as it expands and cools, known as shock cooling emission. Interacting core-collapse supernovae (Types IIn and Ibn) are modeled using \pkg{Arnett\_CSM} \citep{CSM_interaction_model}, which adds a circumstellar-material shock breakout and cooling component. Kilonova light curves for both BNS and NSBH systems are simulated using the same model adopted in ELAsTiCC \citep{ellastic, bulla2019, Lukosiute_2022}, which is a different
model than the prior predictive distribution adopted in \pkg{KilonovaSCORER}. Each model class is simulated 100 times with independent realizations of the model parameters. All simulations follow a Rubin/LSST silver ToO strategy \citep{too_Andreoni2022}, with $g$- and $r$-band imaging on nights 1--4, per-band limiting
magnitudes set to the LSST reference cadence, and only detections with $\mathrm{SNR} > 3$ retained. The luminosity distance is fixed to the value derived from S250818k, and supernovae are allowed to explode up to 20\,d prior to the gravitational-wave trigger to reproduce the range of rest-frame phases most likely to produce kilonova-like photometric signatures at first detection.

\begin{figure*}[!htbp]
    \centering
    \includegraphics[width=\textwidth]{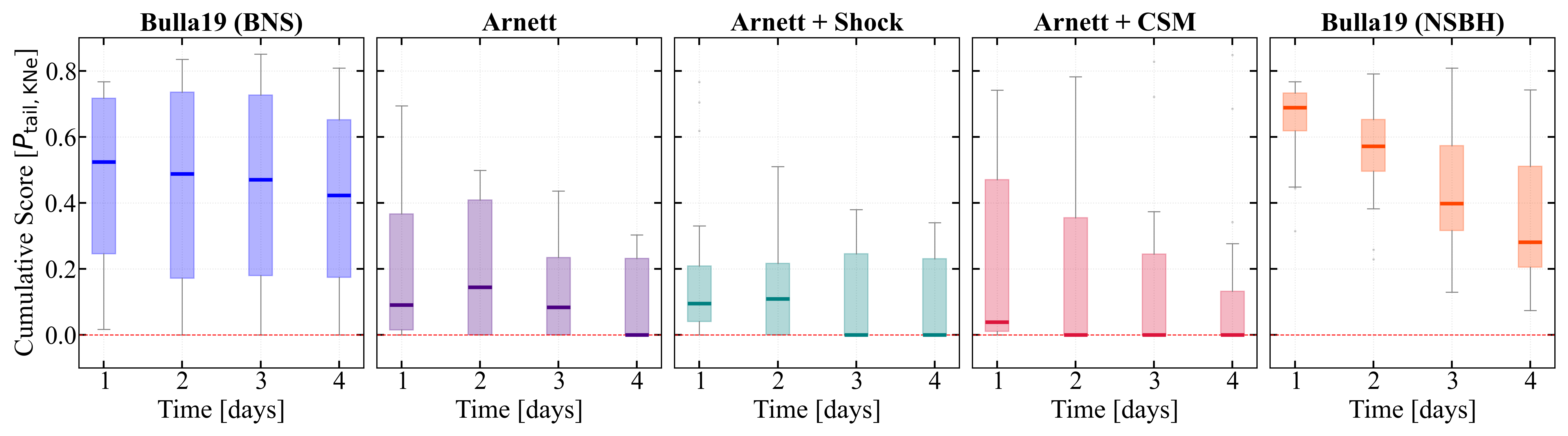}
    \caption{Distribution of cumulative $P_{\mathrm{tail},\mathrm{KNe}}$
    scores as a function of time after the gravitational-wave trigger for
    five simulated transient classes: BNS kilonova (Bulla19 BNS, blue),
    Type Ia supernova (Arnett, purple), core-collapse supernova with shock
    cooling (Arnett + Shock, teal), interacting core-collapse supernova
    (Arnett + CSM, red), and NSBH kilonova (Bulla19 NSBH, orange). Each
    box shows the interquartile range (IQR) of the cumulative score across
    100 independent simulations at $t = 1$, 2, 3, and 4\,d post-trigger. The horizontal
    line within each box is the median, and whiskers extend to the adjacent values within $1.5\times$ IQR. Horizontal axis: days since gravitational-wave trigger. Vertical axis: cumulative $P_{\mathrm{tail},\mathrm{KNe}}$.}
    \label{fig:all_sim}
\end{figure*}

Figure~\ref{fig:all_sim} shows the distribution of cumulative
$P_{\mathrm{tail},\mathrm{KNe}}$ scores as a function of time after the
gravitational-wave trigger for all simulated transient classes. Both
kilonova populations maintain consistently high scores throughout the
four-day observing window. BNS mergers achieve a median cumulative score
of $0.52^{+0.19}_{-0.28}$ at $t = 1$\,d, and remain relatively stable at
$\sim 0.42$ through the end of the observing plan. NSBH mergers start
with a higher median score of $0.68^{+0.04}_{-0.07}$ at $t = 1$\,d but
show a steeper decline, consistent with the faster post-peak evolution
characteristic of NSBH kilonova light curves relative to BNS systems. In
both cases the score distributions remain above the zero threshold at all epochs.

All three supernova classes are discriminated from the kilonova
populations within four days of the GW trigger. Arnett +
CSM supernovae (Types IIn, Ibn) show the most rapid suppression, with a
median score of $0.04^{+0.43}_{-0.03}$ at $t = 1$\,d and a median
collapse to zero by $t = 2$\,d. Arnett + shock cooling
supernovae (Types II, IIb) follow a similar trajectory, with a median
score of $0.10^{+0.11}_{-0.05}$ at $t = 1$\,d collapsing to zero by
$t = 3$--$4$\,d. Type Ia supernovae exhibit the most kilonova-like
behavior at early times, with the highest median cumulative score of
$0.11 \pm 0.10$ at $t = 2$\,d, reflecting the photometric degeneracy
between thermonuclear supernovae and kilonovae near the expected kilonova
peak. Despite this, the median Type Ia score collapses to zero by
$t = 4$\,d. By the final observed epoch, \pkg{KilonovaSCORER} ranks simulated BNS and NSBH kilonovae higher than the supernova contaminants, especially after three days, when all supernova populations have
median scores below $0.1$. 

\section{Discussion}\label{sec:discussion}

\subsection{Broad Implications}

\begin{table*}[!t]
\centering
\caption{Qualitative comparison of approaches for early-time kilonova candidate assessment in alert-driven follow-up. ``Latency'' refers to typical computational and decision-making time per candidate, not broker ingest delay.}
\label{tab:comparison_methods}
\begin{tabular}{lccc}
\hline\hline
Approach & Data/Input requirements & Latency & Role of kilonova physics \\
\hline
Full Bayesian inference & Multi-epoch light curves, often dense & High & Explicit likelihood and priors \\
ML taxonomic classifiers & Engineered features from alert streams & Low & Implicit in training datasets \\
\pkg{KilonovaSCORER} & Sparse photometry ($m,\sigma,t$ bands) and GW distance & Moderate & Explicit prior predictive grid \\
\hline
\end{tabular}
\end{table*}

\pkg{KilonovaSCORER} is designed to operate in real time as
photometric observations are ingested during gravitational-wave
follow-up campaigns, making it directly applicable to the high-volume
transient stream expected from Rubin/LSST (see Table~\ref{tab:comparison_methods}), and additional data from wide-field facilities like LS4 \citep{LS4_Miller2025} and the Zwicky Transient Facility (ZTF; \cite{ZTF_Dekany2020}). As the number of
gravitational-wave events and their associated optical candidates grows
with improved detector sensitivity, automated quantitative scoring
tools become essential for prioritizing spectroscopic and
multi-wavelength follow-up resources. The framework is agnostic to
the gravitational-wave source classification: even for events without
a clear BNS or NSBH distinction, \pkg{KilonovaSCORER} can be
applied directly, as demonstrated in Section~\ref{sec:othertransients}.
It is intended as one component within a broader ecosystem of
real-time multimessenger analysis tools, and its output integrates
naturally into follow-up coordination platforms, alert brokers and Target and Observation Managers \citep{TOM_toolkit}, such as \cite{Coughlin2023_skyportal,Coulter23,AI_scope}.
The per-epoch scores and survival fraction $f_{\mathrm{surv}}(t)$ can
also serve as physics-informed metadata features for early-time
machine learning classifiers, providing model-grounded information
beyond standard light-curve summary statistics.

\subsection{Caveats}

Several scenarios can lead to supernovae or other non-kilonova transients maintaining high ranks within this framework over limited observational baselines. This occurs when the sampled phase of the light curve is photometrically degenerate with kilonova evolution, either because observations terminate before divergence becomes apparent (e.g., SN\,2025ulz at $t \lesssim 2.5$\,d), or because the transient exploded before the GW trigger and only a portion of its declining tail is observed. In such cases, the sequential ABC filter will ultimately suppress the cumulative score to zero once sufficient multi-epoch data are available. Accordingly, the framework should not be interpreted as a standalone rejection criterion in low-cadence regimes, but rather as a dynamic ranking metric that provides early-time diagnostic power, even from sparse photometric sampling.

A further limitation is that scores are always conditioned on the adopted simulation grid: candidates whose emission physics falls outside that prior (for example, different heating or opacity prescriptions) may be mis-ranked relative to grid-compatible kilonovae, even when the transient is astrophysical. 

\subsection{Extensions}

\textbf{GRB-associated kilonovae:}
For kilonovae associated with GRB events, the merger time
is known from the prompt emission but the luminosity distance may
carry significant uncertainty depending on whether a host galaxy
redshift is available. In this regime \pkg{KilonovaSCORER} can be
adapted to use either a host galaxy distance or a broad distance
prior, with the merger time fixed to the GRB trigger and the kilonova
model grid extended to include an afterglow component. If the
afterglow is independently constrained from radio and $\gamma$-ray data, the
scoring framework can be applied in its default configuration.

\textbf{Untargeted kilonova searches:}
In the absence of a GW or GRB trigger, for example,
in a photometric search for fast red transients during the wide-field LSST survey, both
the merger time and the luminosity distance are unconstrained.
\pkg{KilonovaSCORER} can be adapted to this regime by marginalizing
over a grid of merger times and adopting a host galaxy
photometric redshift or a broad luminosity distance prior, with a
wider time bin to account for the merger time uncertainty at the cost of
reduced discriminating power between consecutive epochs. The key idea
is that for a single observation there always exists a $t_0$ that
maximizes the score, and as additional observations become available, the
optimal $t_0$ is that which simultaneously maximizes consistency
across all epochs. This approach requires a larger simulation grid
and longer scoring time, but provides a principled vetting procedure
for candidates whose intrinsic brightness is incompatible with kilonova
expectations at any assumed merger time, particularly for overluminous
classes such as Luminous Fast Blue Optical Transients
\citep[LFBOTs,][]{anya_LFBOTs} and Superluminous Supernovae
\citep[SLSNe,][]{Nicholl2017_Magnetar_SLSN}.

\textbf{Binary black hole mergers in AGN disks:}
The framework extends naturally to electromagnetic counterpart
candidates associated with binary black hole mergers in AGN accretion
disks \citep{Darc2025}. In this case, the adopted model grid describes radiation mechanisms such as jet breakout emission, disk cocoon cooling, and jet cocoon cooling \citep{Chen_2024}, but the scoring and ABC diagnostic
logic are unchanged. More broadly, \pkg{KilonovaSCORER} is
applicable to any astrophysical transient class for which a prior
predictive distribution of absolute magnitudes can be constructed,
provided that the explosion time and a distance estimate are available for each candidate.

\vspace{0.5em}
\noindent
Finally, it is important to emphasize that \pkg{KilonovaSCORER} is
a ranking tool, not a classifier. It does not provide
a fixed decision threshold for unequivocally accepting or rejecting
candidates, and should not be used as a standalone rejection
criterion. Its primary purpose is to deliver quantitative,
interpretable metrics and a visual diagnostic report to support
rapid, data-driven decisions during follow-up campaigns operating
under strict time and resource constraints.

\section{Conclusion}
\label{sec:conclusion}

We have presented \pkg{KilonovaSCORER}, an open-source, real-time
photometric scoring framework for ranking kilonova candidates during
gravitational-wave and multimessenger follow-up campaigns. The key
results and contributions of this work are summarized as follows.

\begin{itemize}

    \item We introduced two complementary per-observation metrics,
    $P_{\mathrm{tail},\mathrm{KNe}}$ and $P_{\mathrm{near},\mathrm{KNe}}$,
    grounded in prior predictive ideas from the statistical literature on simulation-based methods. $P_{\mathrm{tail},\mathrm{KNe}}$ measures how extreme each observation is in the marginal prior predictive distribution at $(t,b)$ (a local tail probability), while $P_{\mathrm{near},\mathrm{KNe}}$ measures the fraction of prior samples whose predicted magnitudes fall inside the observation ROPE (a local ``near'' probability). Global temporal consistency across epochs is enforced separately by the ABC survival diagnostic, not by these two metrics alone.

    \item The cumulative $P_{\mathrm{tail},\mathrm{KNe}}$ score
    aggregates per-observation scores across time bins using an
    inverse-variance weighted mean (IVWM) in logit space, producing a final
    score in $[0, 1]$. This score is used to rank
    candidates within a GW search in a statistically
    principled way that accounts for both the number and quality of
    available photometric measurements.

    \item We introduced an ABC-inspired sequential survival diagnostic
    that tracks the fraction of simulated light curves globally
    consistent with the cumulative photometric history of a candidate.
    This diagnostic provides a physically motivated penalization
    mechanism that sets the cumulative score to zero as soon as the
    observed evolution becomes incompatible with any model in the
    kilonova grid, regardless of individual epoch scores.

    \item Applied to AT\,2017gfo (the only confirmed kilonova with an
    associated GW detection), \pkg{KilonovaSCORER} recovers
    consistently high cumulative scores across all observed bands and
    epochs, demonstrating that the framework correctly identifies and
    ranks a genuine kilonova with high confidence.

    \item Applied to SN\,2025ulz, a Type IIb supernova initially
    proposed as a kilonova counterpart to S250818k, the framework
    assigns a non-negligible early score consistent with its
    kilonova-like photometric appearance at $t \lesssim 2.5$\,d, but
    the ABC diagnostic collapses the cumulative score to zero once the
    re-brightening at $t \approx 5$\,d is incorporated, correctly
    ruling out the candidate without requiring spectroscopic
    classification.

    \item In a Rubin/LSST ToO simulation spanning three
    supernova emission models (Arnett, Arnett + shock cooling, Arnett + CSM)
    and two kilonova populations (BNS and NSBH), all
    supernova classes are discriminated from kilonovae within four
    days of the gravitational-wave trigger on average. Kilonova populations maintain median cumulative scores of ${\sim}\,0.6$--$0.4$ throughout the
    observing window, while all supernova median scores collapse to
    zero by $t = 3$--$4$\,d.

\end{itemize}

\section{Software and Data Availability}
\label{sec:software_data}

The \pkg{KilonovaSCORER} pipeline described in this paper is publicly available at \url{https://github.com/phelipedarc/KilonovaSCORER/tree/main}.
Benchmark light curves for AT\,2017gfo, SN\,2025ulz, and the GRB-hosted kilonova sample are taken from the literature as cited in Section~\ref{sec:results}.
Simulated transient populations used for the ranking tests in Section~\ref{sec:othertransients} are generated with \pkg{redback} as summarized there.

\begin{acknowledgments}
We thank the LIGO--Virgo--KAGRA Collaboration for public gravitational-wave data products and the teams behind the Open Supernova Catalog, the Transient Name Server, and the General Coordinates Network for open transient data access. P.D.\ acknowledges support from the Artificial Intelligence for Physics Laboratory (Lab-IA) and the Centro Brasileiro de Pesquisas F\'{i}sicas (CBPF), and thanks Rafael S.\ de Souza for valuable discussions on statistical analysis. C.D.K.\ gratefully acknowledges support from the NSF through AST-2432037, the HST Guest Observer Program through HST-SNAP-17070 and HST-GO-17706, and from JWST Archival Research through JWST-AR-6241 and JWST-AR-5441. This research has made use of NASA's Astrophysics Data System Bibliographic Services.
\end{acknowledgments}





%
\facilities{LIGO, Rubin:SimonyiSurveyTelescope, Zwicky Transient Facility}

\software{\pkg{KilonovaSCORER} (\url{https://github.com/phelipedarc/KilonovaSCORER/tree/main}),
          \pkg{redback} \citep{REDBACKSarin2024},
          \pkg{Astropy} \citep{2013A&A...558A..33A,2018AJ....156..123A,2022ApJ...935..167A},
          \pkg{NumPy} \citep{numpy2020}, \pkg{SciPy} \citep{scipy2020}
          }

\bibliography{PASPsample701}{}
\bibliographystyle{aasjournalv7}

\appendix

\section{\texorpdfstring{\pkg{KilonovaSCORER}: Metric and Simulation Settings}{KilonovaSCORER: Metric and Simulation Settings}}
\label{app:settings}

\subsection*{Metric Choice}

The framework supports different choices of acceptance criterion within
the ABC diagnostic and different scoring metrics. For example, instead
of the Region of Practical Equivalence, one may adopt alternative
distance measures such as the mean squared error or mean absolute
error with an acceptance threshold $\varepsilon$. If the threshold is
defined as $\varepsilon = k\,\sigma_{\mathrm{obs}}$, such that a
simulation is accepted whenever
\begin{equation}
    \left| M_{\mathrm{sim}} - M_{\mathrm{obs}} \right| < k\,\sigma_{\mathrm{obs}},
\end{equation}
this condition is equivalent to the ROPE criterion adopted in this
work, where acceptance is defined relative to a tolerance region
determined by the observational uncertainty.

A likelihood-based scoring function is also available. Assuming
Gaussian observational uncertainties, the log-likelihood is
\begin{equation}
    \log \mathcal{L} = -\frac{1}{2} \sum_i \left[
    \frac{\left( M_{\mathrm{sim},i} - M_{\mathrm{obs},i} \right)^2}
    {\sigma_{\mathrm{obs},i}^2}
    + \log\left( 2\pi\,\sigma_{\mathrm{obs},i}^2 \right) \right],
\end{equation}
which can be used either as a scoring function or as a statistical
diagnostic. In this work we adopt the ROPE-based diagnostics
described in Section~\ref{sec:kilonovascorer} because they are better suited
to the expected properties of gravitational-wave follow-up data:
sparse observations, limited per-epoch statistics, and simulated
absolute magnitude distributions that are often non-Gaussian and
occasionally multimodal.

\begin{figure}[!htbp]
    \centering
    \includegraphics[width=0.7\textwidth]{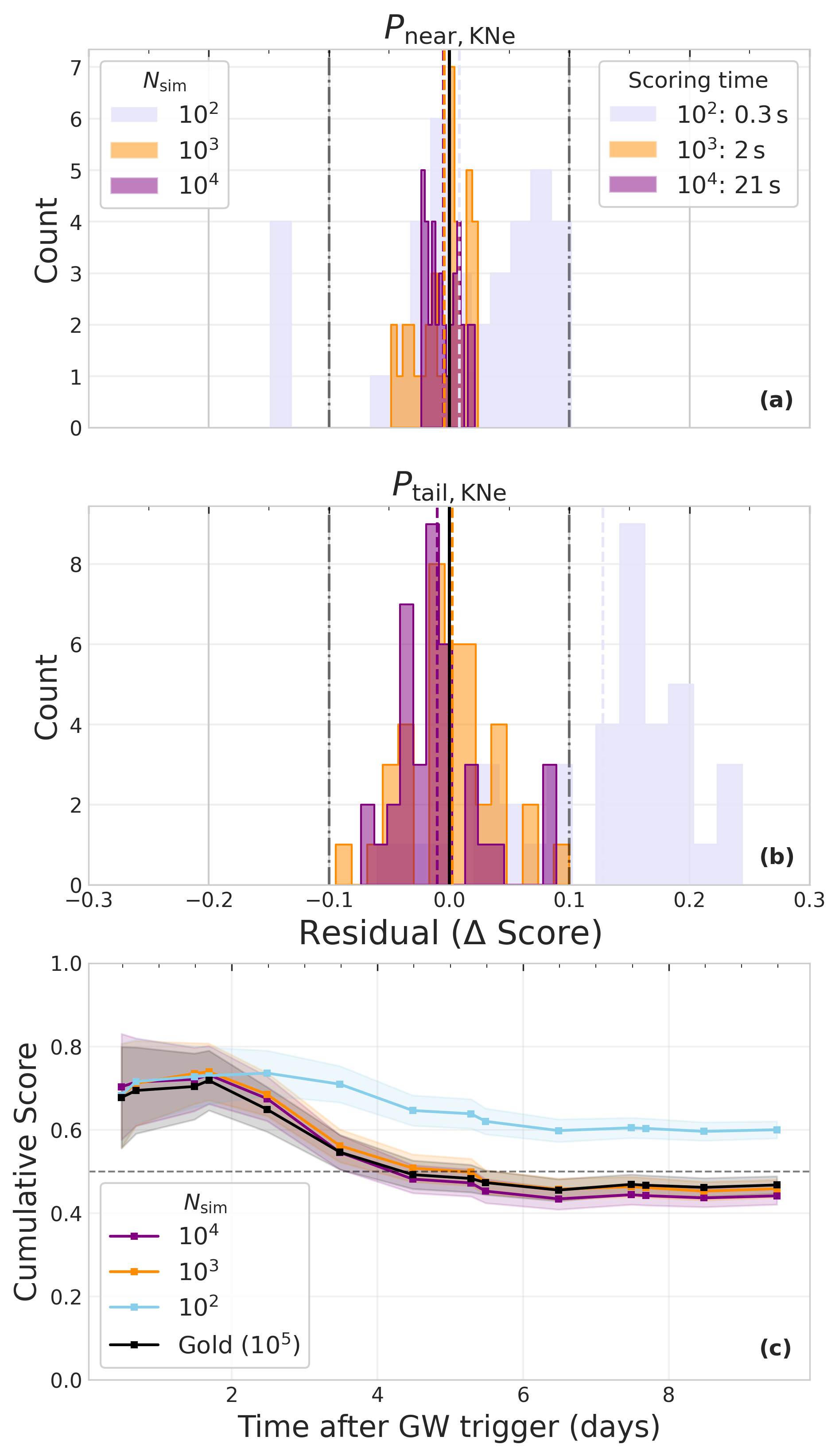}
    \caption{Sensitivity of \pkg{KilonovaSCORER} to the size of
    the simulation grid, evaluated on 38 observations of AT\,2017gfo
    across the $g$-, $r$-, $i$-, and $z$-bands. \textit{Left panels:}
    residuals $\Delta P_{\mathrm{near},\mathrm{KNe}}$ (top) and
    $\Delta P_{\mathrm{tail},\mathrm{KNe}}$ (bottom) relative to the
    gold-standard scores ($N = 10^5$), shown as a function of epoch
    for the high ($N = 10^4$, orange), medium ($N = 10^3$, green),
    and low ($N = 10^2$, red) simulation sets. The grey band indicates
    the characteristic $1\sigma$ uncertainty of the gold-standard
    $P_{\mathrm{tail},\mathrm{KNe}}$ score. \textit{Right panel:}
    wall-clock scoring time in seconds as a function of simulation set
    size, demonstrating a ${\sim}100\times$ reduction in compute time
    when using the medium set relative to the gold standard with no
    significant loss in score accuracy.}
    \label{fig:size_data}
\end{figure}

\subsection*{Simulation Choice}

We explored several alternative simulation setups before adopting our
fiducial two-component kilonova model. A single-component model with
broader priors based on \citet{Metzger17} was tested first, but most
prior samples produced red, rapidly declining light curves that
yielded systematically low scores for AT\,2017gfo at epochs later than
${\sim}6$\,d. We also tested a viewing-angle-dependent model based on
the radiative transfer simulations of \citet{bulla2019}, as adopted in
the \textit{ELAsTiCC} dataset. While this model provided good
agreement with AT\,2017gfo, the resulting light curves exhibited
relatively small magnitude dispersion across realizations, particularly
during the first few days after merger. This limited diversity reduces
robustness for early-time candidate scoring. Since our primary goal is
to avoid prematurely down-ranking genuine kilonova candidates during the
early phase of evolution, we adopted a two-component model with
exploration of the extreme regions of its parameter space, providing
broader coverage of plausible kilonova behaviors at early times.

\subsection*{Sensitivity to the Number of Simulations}

To assess the sensitivity of \pkg{KilonovaSCORER} and the ABC
diagnostic to the size of the simulation grid, we evaluated the
scoring pipeline on four sets of increasing size: a gold-standard set
of $N = 10^5$ simulations and three reduced sets comprising $N = 10^2$
(low), $10^3$ (medium), and $10^4$ (high) draws, all sampled from the
same prior. As a test case, we scored 38 individual observations of
AT\,2017gfo across the $g$-, $r$-, $i$-, and $z$-bands, retaining the
highest signal-to-noise ratio observation per band per day.

Figure~\ref{fig:size_data} shows the residuals $\Delta
P_{\mathrm{near},\mathrm{KNe}}$ and $\Delta P_{\mathrm{tail},\mathrm{KNe}}$
relative to the gold-standard scores, together with the 
time required to score all observations in seconds. Both the high ($10^4$) and
medium ($10^3$) sets yield residuals within the characteristic
uncertainty of $P_{\mathrm{tail},\mathrm{KNe}}$
($|\Delta| \lesssim 0.1$) across all epochs and bands, with
$\Delta P_{\mathrm{near},\mathrm{KNe}}$ dispersions of order $0.05$.
The low set ($10^2$) exhibits significantly larger scatter and a
systematic tendency to overestimate scores, particularly within the
first few days post-merger. These results demonstrate that
\pkg{KilonovaSCORER} can be accelerated by a factor of
${\sim}100\times$ using the medium simulation set without
compromising score reliability.

The number of simulations also critically affects the ABC acceptance
threshold. As the simulation grid shrinks, fewer draws satisfy the
ROPE criterion, and the accepted subset may collapse to zero for
aggressive choices of the overlap parameter $k_{ABC}$. For AT\,2017gfo, the
minimum viable threshold is $k_{ABC,\min} = 1.5$ for the gold-standard
set, rising to $k_{ABC,\min} = 2.0$, $2.5$, and $2.0$ for the high,
medium, and low sets respectively. Users should therefore calibrate
$k_{ABC}$ to the size of their available simulation grid. The acceptance
collapse in AT2017gfo typically occurs between 4.5 and 6\,d post-merger, driven
by a slight excess in the $r$-band flux relative to the simulated
light curves.

\section{Kilonovae Associated to GRB Events: Diagnostic Report}
\label{sec:grb_appendix}

\begin{figure}[h!]
    \centering
    \includegraphics[width=0.7\linewidth]{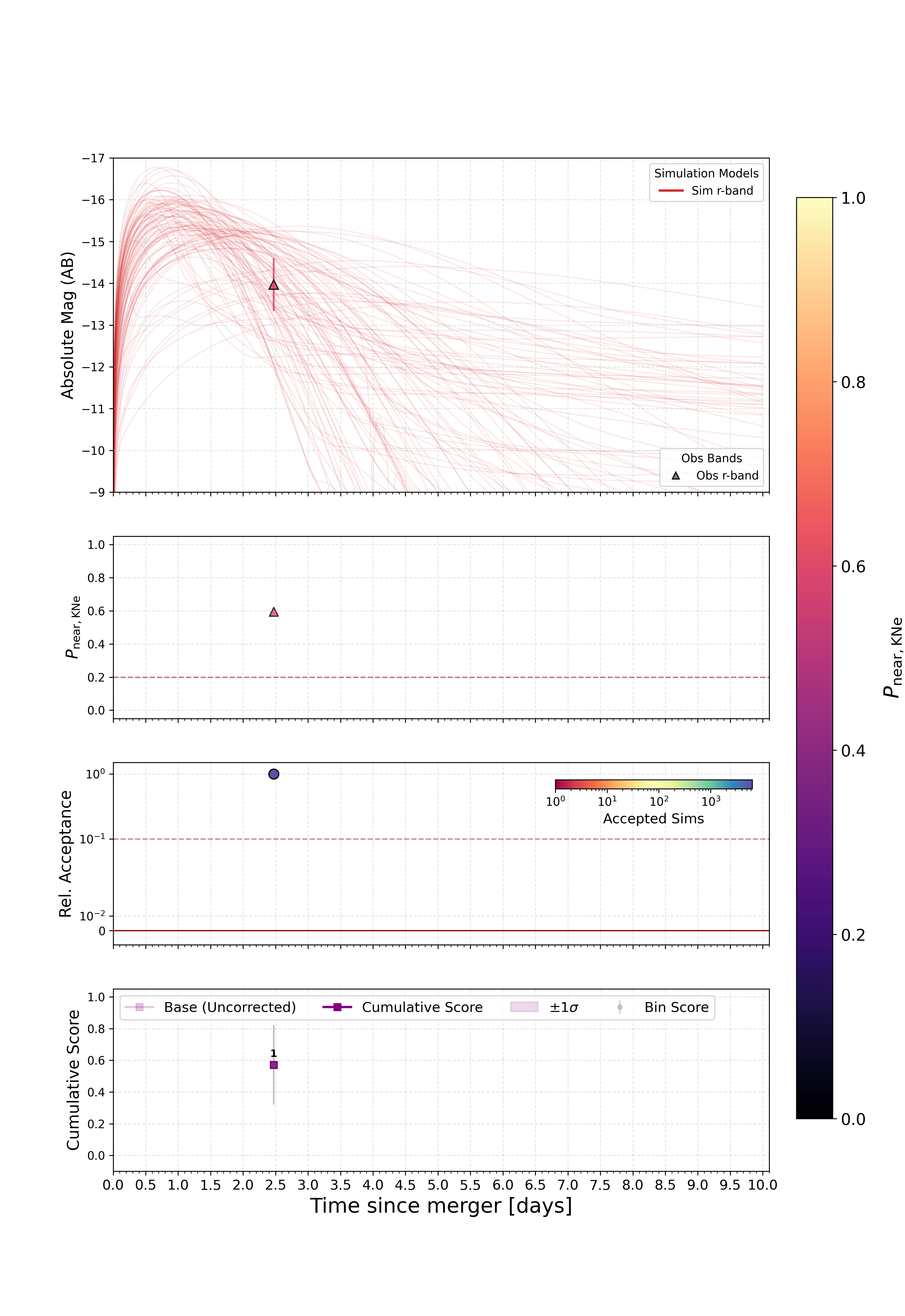}
    \caption{\textbf{GRB050709:}Candidate Diagnostic Report for the afterglow subtracted kilonova observations scored in the $griz$ + J-, H-, K- F356W, F444W bands. }
    \label{fig:GRB050709}
\end{figure}

\begin{figure}[h!]
    \centering
    \includegraphics[width=0.7\linewidth]{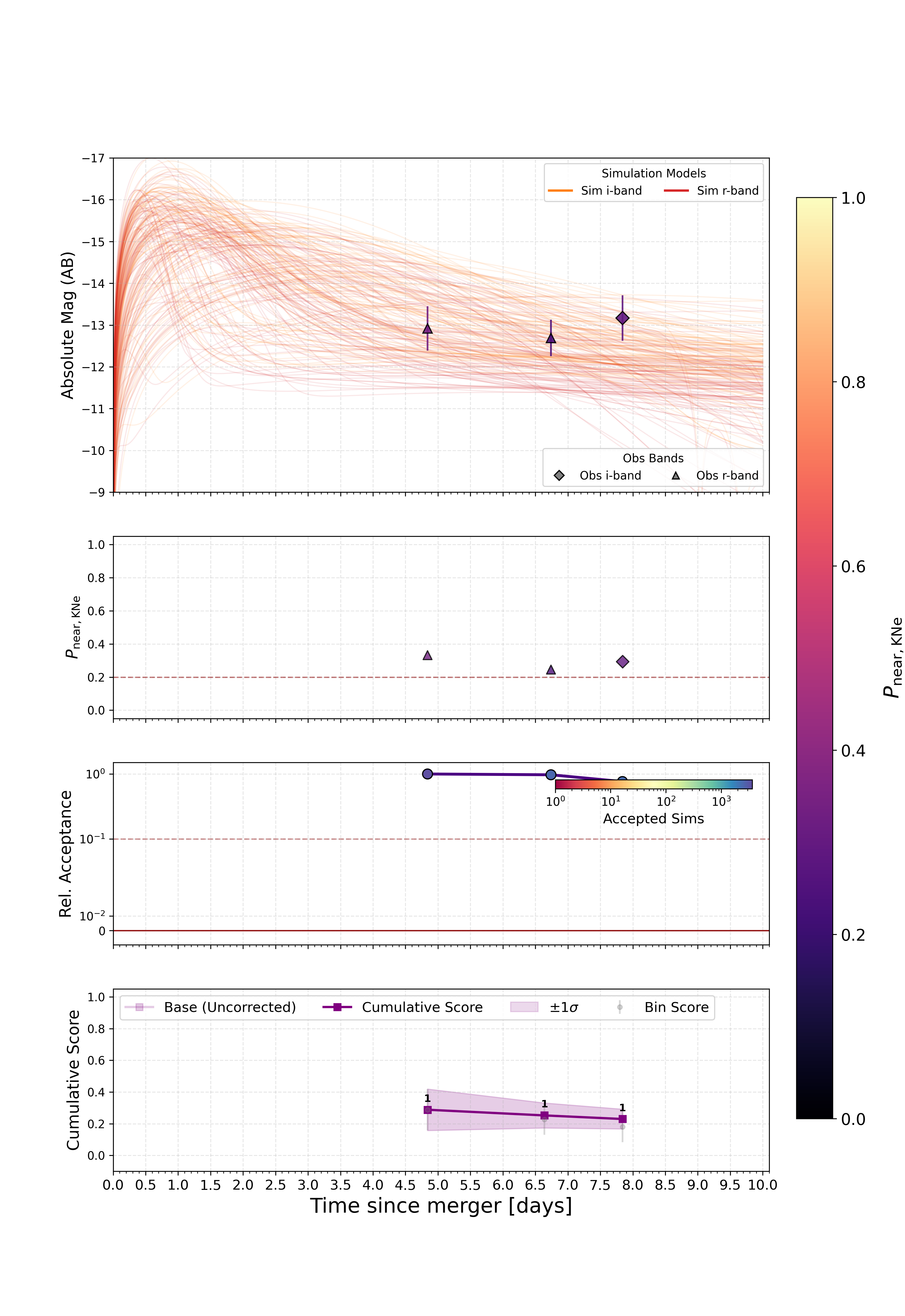}
    \caption{\textbf{GRB060614:} Candidate Diagnostic Report for the afterglow subtracted kilonova observations scored in the $griz$ + J-, H-, K- F356W, F444W bands. }
    \label{fig:GRB060614}
\end{figure}

\begin{figure}[h!]
    \centering
    \includegraphics[width=0.7\linewidth]{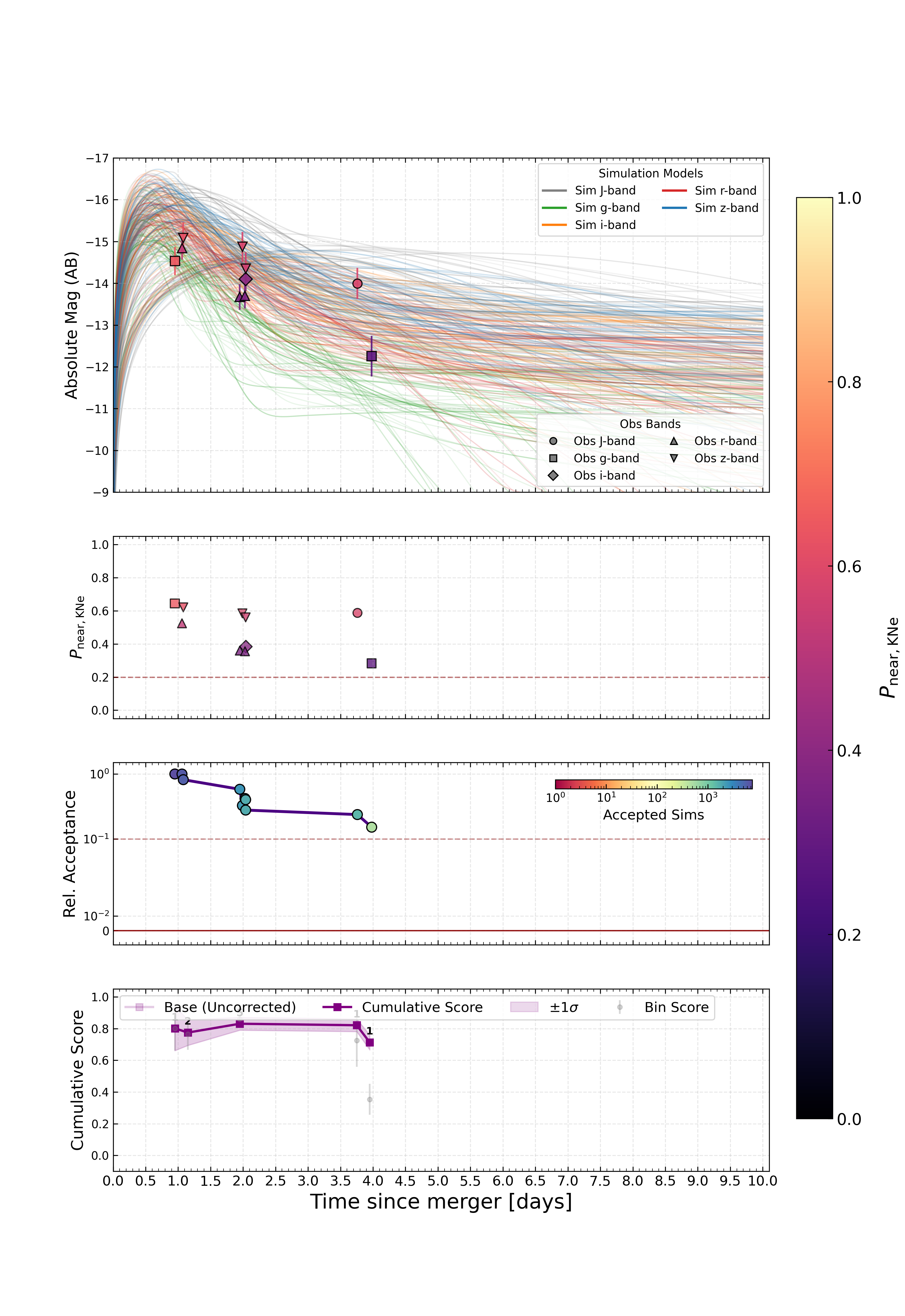}
    \caption{\textbf{GRB160821B:} Candidate Diagnostic Report for the afterglow subtracted kilonova observations scored in the $griz$ + J-, H-, K- F356W, F444W bands. }
    \label{fig:GRB160821B}
\end{figure}

\begin{figure}[h!]
    \centering
    \includegraphics[width=0.7\linewidth]{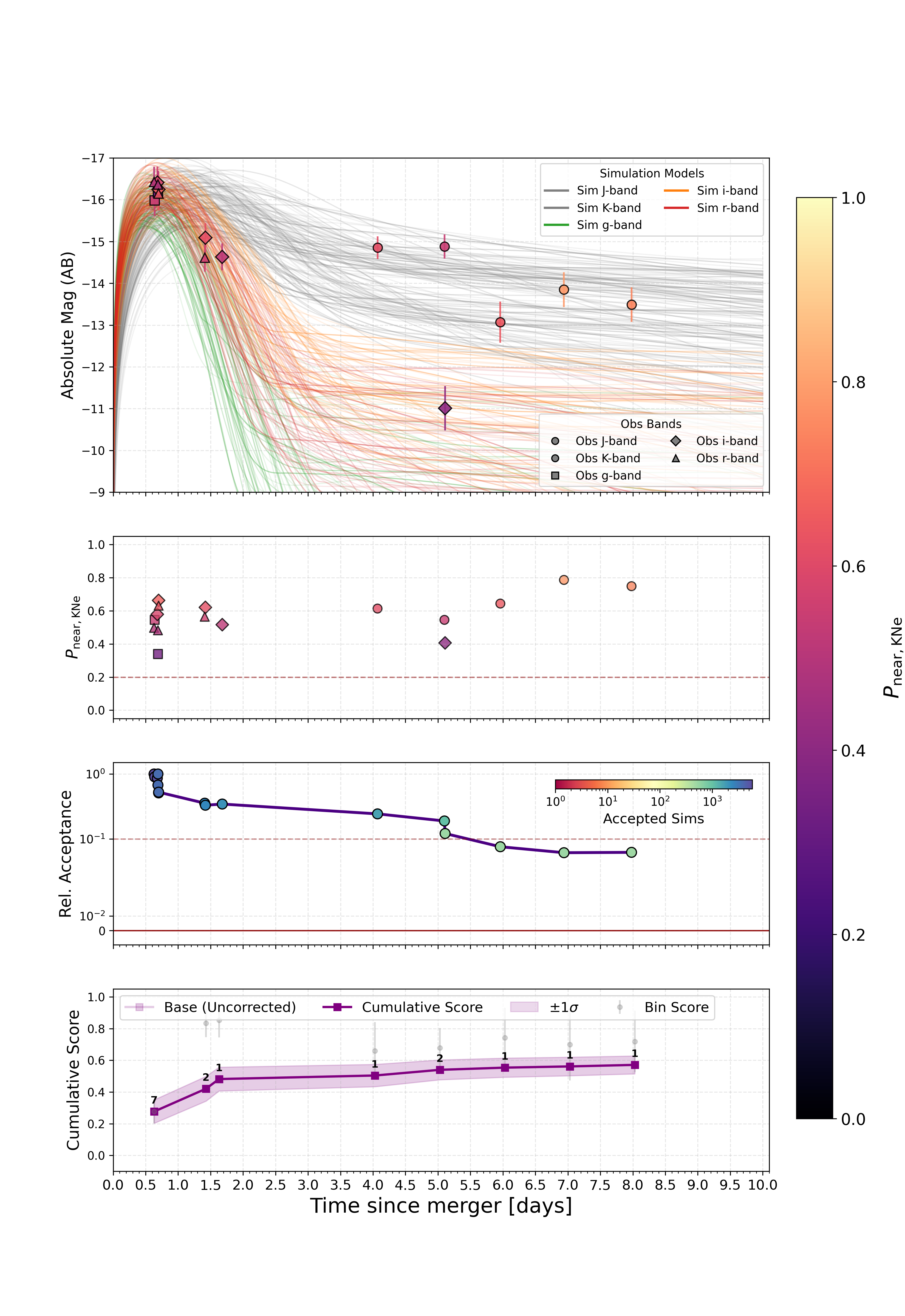}
    \caption{\textbf{GRB211211A:} Candidate Diagnostic Report for the afterglow subtracted kilonova observations scored in the $griz$ + J-, H-, K- F356W, F444W bands. }
    \label{fig:GRB211211A}
\end{figure}

\begin{figure}[h!]
    \centering
    \includegraphics[width=0.7\linewidth]{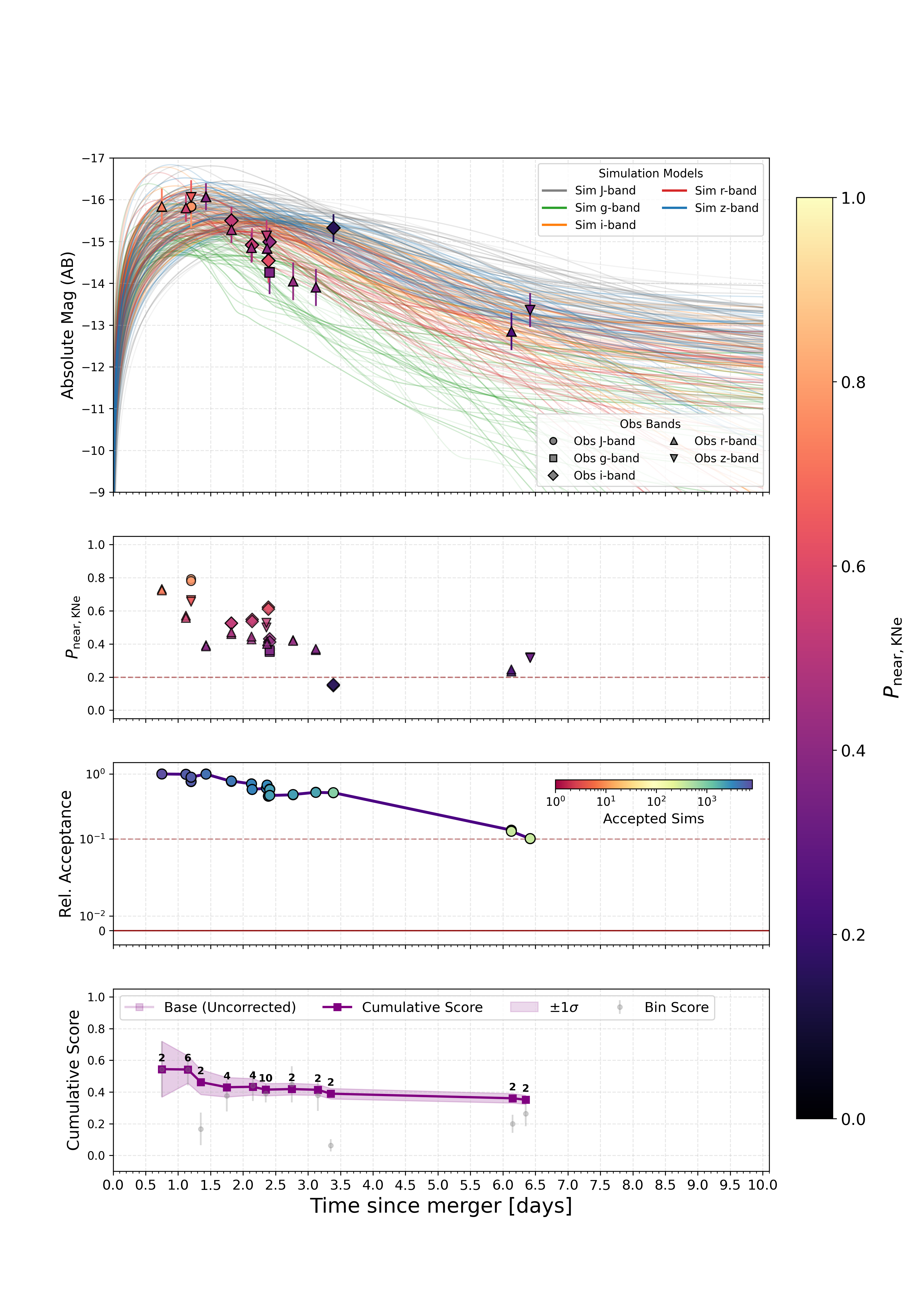}
    \caption{\textbf{GRB230307A:} Candidate Diagnostic Report for the afterglow subtracted kilonova observations scored in the $griz$ + J-, H-, K- F356W, F444W bands. }
    \label{fig:GRB230307A}
\end{figure}

\end{document}